\documentclass[twocolumn,prb]{revtex4}

\usepackage{graphicx}
\usepackage{epsfig}
\usepackage{amsmath, amssymb}
\usepackage{verbatim}

\begin{document}

\title{$U(1) \times U(1) \rtimes Z_2$ Chern-Simons Theory \\
and $Z_4$ Parafermion Fractional Quantum Hall States}

\author{Maissam Barkeshli}
\author{Xiao-Gang Wen}
\affiliation{Department of Physics, Massachusetts Institute of Technology,
Cambridge, MA 02139, USA }

\begin{abstract}

We study $U(1) \times U(1) \rtimes Z_2$ Chern-Simons theory with
integral coupling constants $(k,l)$ and its relation to certain
non-Abelian fractional quantum Hall (FQH) states.  For the $U(1)
\times U(1) \rtimes Z_2$ Chern-Simons theory, we show how to compute
the dimension of its Hilbert space on genus $g$ surfaces and how this
yields the quantum dimensions of topologically distinct excitations.
We find that $Z_2$ vortices in the $U(1) \times U(1)\rtimes Z_2$
Chern-Simons theory carry non-Abelian statistics and we show how to
compute the dimension of the Hilbert space in the presence of $n$
pairs of $Z_2$ vortices on a sphere.  These results allow us to show
that $l=3$ $U(1)\times U(1) \rtimes Z_2$ Chern-Simons theory is the
low energy effective theory for the $Z_4$ parafermion (Read-Rezayi)
fractional quantum Hall states, which occur at filling fraction $\nu =
\frac{2}{2k - 3}$.  The $U(1)\times U(1) \rtimes Z_2$ theory is more
useful than an alternative $SU(2)_4\times U(1)/ U(1)$ Chern-Simons
theory because the fields are more closely related to physical degrees
of freedom of the electron fluid and to an Abelian bilayer phase on
the other side of a two-component to single-component quantum phase
transition. We discuss the possibility of using this theory to
understand further phase transitions in FQH systems, especially the
$\nu = 2/3$ phase diagram.

\end{abstract}

\maketitle

\section{Introduction}

One of the most exciting breakthroughs in condensed matter physics has
been the discovery that there exist quantum phases of matter at zero
temperature that cannot be described by their pattern of symmetry
breaking.\cite{Wtop} The prototypical and perhaps most well-studied
examples of these phases are the fractional quantum Hall
states,\cite{TSG8259} which exhibit a different kind of order, called
topological order.\cite{Wrig} Topologically ordered phases are
currently the subject of intense interest because of the possibility
of detecting, for the first time, excitations that exhibit non-Abelian
statistics,\cite{MR9162,Wnab} and subsequently manipulating these
non-Abelian excitations for robust quantum information storage and
processing.\cite{K032,FKL0331,DKL0252} 

One way to improve our understanding of topological order in the
fractional quantum Hall states is to study phase transitions between
states with different topological order. While much is known about
phase transitions between phases with different patterns of symmetry
breaking, much less is known about phase transitions between phases
with different topological order. Aside from its intrinsic interest,
such information may be useful in identifying the topological order of
a certain FQH state, which is currently a significant challenge. The
experimental observation of a continuous phase transition in a FQH
system may help us identify the topological order of one of the phases
if we know theoretically which topologically ordered phases can be
connected to each other through a continuous phase transition and
which cannot. Ultimately, we would like to have an understanding of
all of the possible topological orders in FQH states and how they can
be related to each other through continuous phase transitions. 

We may hope to understand a phase transition between two phases if we
have a field theory that describes each phase and we know how the
field theories of the two phases are related to each other.  In the
case of the fractional quantum Hall states, it is well-known that the
long-distance, low energy behavior is described by certain topological
field theories in 2+1 dimensions,\cite{BW9045} called Chern-Simons
theories. For the Laughlin states and other Abelian FQH states, such
as the Halperin states, the hierarchy states, and Jain states, the long
wavelength behavior is described by Chern-Simons theories with a
number of $U(1)$ gauge fields.\cite{BW9045,Wtoprev,ZH8982} 

For the non-Abelian FQH states, the corresponding Chern-Simons 
theory has a non-Abelian gauge group.\cite{FN9804,Wpcon} The most
well-studied examples of non-Abelian FQH states are the Moore-Read
Pfaffian state\cite{MR9162} and some of its generalizations, the Read-Rezayi (or
$Z_k$ parafermion) states.\cite{RR9984} The bosonic $\nu=1$ Pfaffian is described
by $SU(2)_2 $ Chern-Simons theory,\cite{FN9804} or alternatively, by
$SO(5)_1 $ Chern-Simons theory,\cite{Wpcon} while the effective 
theories for the other states are less well-understood. 
It has been proposed that the Read-Rezayi $Z_k$ parafermion states 
are described by $SU(2)_k \times U(1)/U(1)$ Chern-Simons theory.\cite{CF0057}

In this paper, we show that Chern-Simons theory with gauge group $U(1)
\times U(1) \rtimes Z_2$ describes the long-wavelength properties of
the $Z_4$ parafermion Read-Rezayi FQH state. The significance of this
result is that there is a bilayer state, the $(k, k, k - 3)$ Halperin
state at $\nu = \frac{2}{2k-3}$, which may undergo a bilayer to
single-layer quantum phase transition to the $Z_4$ parafermion state
as the interlayer tunneling is increased.\cite{RWR} The bilayer phase
is described by a $U(1) \times U(1)$ Chern-Simons theory. This new
formulation of the Chern-Simons theory for the $Z_4$ parafermion state
may therefore be useful in understanding the phase transition because
the gauge groups $U(1) \times U(1) \rtimes Z_2$ and $U(1)\times U(1)$
are closely related, and because the fields in the $U(1) \times U(1)
\rtimes Z_2$ theory are more closely related to physical degrees of
freedom of the electron fluid than they are in the proposed
alternative $SU(2)_4 \times U(1)/U(1)$ theory. 

In addition to aiding us in understanding this phase transition, this
study shows how to compute concretely various topological properties
of a Chern-Simons theory with a disconnected gauge group.  For
Chern-Simons theories at level $k$, where the gauge group is a
simple Lie group $G$, there is a straightfoward prescription to
compute topological properties. The different quasiparticles are
labelled by the integrable highest weight representations of the
affine lie algebra $\hat{g}_k$, where $g$ is the Lie algebra of $G$,
while the quasiparticle fusion rules are given by the Clebsch-Gordon
coefficients of the integrable representations of $\hat{g}_k$.\cite{W8951}  
In contrast, when the gauge group is disconnected, and is of the form $G
\rtimes H$, where $H$ is a discrete automorphism group of $G$, it is
much less straightfoward to compute the topological properties of the
Chern-Simons theory directly. One reason for this is that discrete
gauge theories are most easily studied (and defined) on a lattice,
while it is difficult to formulate lattice versions of Chern-Simons
theories. This complicates the study of Chern-Simons theories with
disconnected gauge groups. 

In the case where the gauge group is $U(1) \times U(1) \rtimes Z_2$,
we show how to compute the ground state degeneracy on genus $g$
surfaces and how this yields the quantum dimensions of the
quasiparticles. We find that the $Z_2$ vortices carry non-Abelian
statistics and we show how to compute the degeneracy of states in the
presence of $n$ pairs of $Z_2$ vortices. The results, for a certain choice
of coupling constants, agree exactly with results obtained in other ways 
for the $Z_4$ parafermion FQH state. 

\section{Motivation and Background}

One interesting way of obtaining the Pfaffian quantum Hall states is
by starting with a bilayer $(k, k, k-2)$ quantum Hall state and taking
the interlayer tunneling to infinity. The bilayer state is at a
filling fraction $\nu = \frac{1}{k-1}$ and is described by the wave
function $\Psi = \Phi(\{z_i\}, \{w_i\}) e^{-\frac{1}{4} \sum_i
(|z_i|^2 + |w_i|^2)}$, with
\begin{equation}
\Phi = \prod_{i<j}^N (z_i-z_j)^k \prod_{i<j}^N (w_i-w_j)^k \prod_{i,j}^N (z_i-w_j)^{k-2} .
\end{equation}
Here, $z_i = x_i + i y_i$ is the complex coordinate of the $i$th
electron in one layer and $w_i$ is the complex coordinate for the
$i$th electron in the other layer. 

As the tunnelling is taken to infinity, we effectively end up with a
single-layer state. The particles in the two layers become
indistinguishable and so we might expect that the resulting
wavefunction is the $(k,k,k-2)$ bilayer wavefunction but
(anti)-symmetrized between the $\{z_i\}$ and $\{w_i\}$ coordinates.
The resulting wavefunction happens to be the Pfaffian state:
\begin{align}
\Psi_{Pf}(\{z_i\}) & = Pf \left( \frac{1}{z_i-z_j} \right) \prod_{i<j}^{2N} (z_i-z_j)^{k-1} 
\nonumber \\
& = S \{\Psi(\{z_i\},\{w_i\}) \},
\end{align}
where $S\{ \cdots \}$ refers to symmetrization or anti-symmetrization
over $z_i$ and $w_i$ depending on whether the particles are bosons are
fermions. Here we have set $z_{N+i} = w_{i}$. Indeed, the $(k, k,
k-2)$ bilayer states undergo a continuous quantum phase transition to
the single-layer $\nu = \frac{1}{k-1}$ Pfaffian states as the
interlayer tunneling is increased.\cite{RG0067,W0050} 

In a similar fashion, the $(k,k,k-3)$ bilayer wave functions, when
(anti)-symmetrized over the coordinates of particles in the two
layers, yield the $Z_4$ parafermion states at filling fraction $\nu =
\frac{2}{2k-3}$.\cite{RWR,RR9984} One way to verify this statement is
through an operator algebra approach that also naturally suggests
$U(1) \times U(1) \rtimes Z_2$ as the appropriate gauge group for the
corresponding Chern-Simons theory (see Appendix
\ref{projConstruction}).  This observation suggests that as the
interlayer tunneling is increased, there may be a region of the phase
diagram where there is a phase transition from the bilayer $(k,k,k-3)$
state to the single-layer non-Abelian $Z_4$ parafermion state.  For $k
= 3$, this is a phase transition at $\nu = 2/3$, the phase diagram of
which has attracted both theoretical and experimental attention. 

Given this perspective, we might expect that we can understand the low
energy effective field theory of the Pfaffian and $Z_4$ parafermion
states by gauging a discrete $Z_2$ symmetry associated with the $Z_2$
symmetry of interchanging the two layers. The effective
field theories for the bilayer states are the $U(1) \times U(1)$
Chern-Simons theories with the field strength of one $U(1)$ gauge field
describing the electron density for  one layer and the field strength of the
other gauge field for the other layer.  This perspective suggests that
the topological properties of these non-Abelian states can be
described by a $U(1)\times U(1) \rtimes Z_2$ Chern-Simons theory. This
is a $U(1) \times U(1)$ Chern-Simons theory with an additional local
$Z_2$ gauge symmetry. The semi-direct product $\rtimes$ here indicates
that the $Z_2$ acts on the group $U(1)\times U(1)$; the $Z_2$ group
element does not commute with elements of $U(1) \times U(1)$. In other
words, elements of the group are $(a, \rho)$, where $a \in U(1) \times
U(1)$ and $\rho \in Z_2$, and multiplication is defined by $(a_1,
\rho_1) * (a_2, \rho_2) = (a_1 \rho_1 a_2 \rho_1, \rho_1 \rho_2)$.
This expectation for $U(1) \times U(1) \rtimes Z_2$ Chern-Simons
theory turns out to be correct for the $Z_4$ parafermion states but
not quite correct for the Pfaffian states, as we will discuss. 

We already have a field theory that correctly describes the
topological properties of the bosonic $\nu=1$ Pfaffian quantum Hall
state. This is the $SU(2)_2$ Chern-Simons theory described in
\Ref{FN9804} or the $SO(5)_1$ Chern-Simons theory described in
\Ref{Wpcon}. (The Pfaffian quantum Hall state at other filling
fractions are described by $SU(2)_2\times U(1)/U(1)$ or $SO(5)_1\times
U(1)/U(1)$ Chern-Simons theory.\cite{Wpcon}) Similarly, the $SU(2)_k
\times U(1)/U(1)$ Chern-Simons theories described in \Ref{CF0057}
encapsulate in some sense the topological properties of the $Z_k$
parafermion states. A possible shortcoming of those theories, however,
is that it can be unclear how to connect the degrees of freedom of the
field theory to the physical degrees of freedom of the electron
liquid.  In contrast, the $U(1)\times U(1) \rtimes Z_2$ makes clearer
the connection between the gauge fields and various physical degrees
of freedom. It also makes clearer the relation to the bilayer state on
the other side of the phase transition. Given this closer contact to
the physical degrees of freedom of the electron fluid and to the
bilayer Abelian phase, it is possible that this point of view may aid
us in understanding physical properties of these quantum Hall states,
such as the quantum phase transition between two topologically ordered
phases: the bilayer Abelian phases and the non-Abelian single-layer
phases. 

The fact that such a Chern-Simons theory might describe the Pfaffian
and/or $Z_4$ parafermion FQH states might also be expected from
another point of view. It is known that the $Z_4$ parafermion
conformal field theory, which is used in constructing the $Z_4$
parafermion FQH states, is dual to the rational $Z_2$ orbifold at a
certain radius.\cite{DV8985} The rational $Z_2$ orbifold at radius $R$
is the theory of a scalar boson $\varphi$ compactified on a circle of
radius $R$, \it i.e. \rm $\varphi \sim \varphi + 2 \pi R$, and that is
gauged by a $Z_2$ action: $\varphi \sim - \varphi$. Furthermore, the
$Z_2$ orbifold at a different radius is dual to two copies of the
Ising CFT, which is used to construct the Pfaffian states. The
Chern-Simons theory corresponding to the $Z_2$ orbifold CFT has gauge
group $O(2)$, which we can think of as $U(1) \rtimes Z_2$.\cite{MS8922} This line
of thinking is what led the authors of \Ref{FH0191} to first
mention that $U(1) \times O(2)$ Chern-Simons theories are related to
the Pfaffian and $Z_4$ parafermion states.  In the $Z_4$ parafermion
case, the relation to $U(1) \times O(2)$ is suggestive but incomplete
because the $U(1)$ and the $O(2)$ need to be ``glued'' together in an
appropriate way; we elaborate more on this point in Appendix \ref{torusDiscussion}.
The proper formulation is the $U(1) \times U(1) \rtimes Z_2$ theory 
that we present here and for which we compute many topological properties. 

Let us first discuss the $U(1) \times U(1)$ Chern-Simons theories that
describe the $(k, k, k-l)$ bilayer states. These are defined by the
Lagrangian 
\begin{equation}
\label{lagrangian}
L = \frac{k}{4\pi}\int_M (a \partial a + \tilde{a} \partial \tilde{a}) + 
\frac{k-l}{4\pi}\int_M (a \partial \tilde{a} + \tilde{a} \partial a),
\end{equation}
where $M$ is a two-dimensional manifold and $a(x,y,t)$ and
$\tilde{a}(x,y,t)$ are two $U(1)$ gauge fields defined on $M \times
\mathbb{R}$. $M$ describes space and $\mathbb{R}$ describes time. The
electron current/density in the top and bottom layers, $j_{\mu}$ and
$\tilde{j}_{\mu}$, respectively, are given by: 
\begin{align} 
j_{\mu} =
\frac{1}{2\pi}\epsilon^{\mu \nu \lambda} \partial_{\nu} a_{\lambda},
\nonumber \\ \tilde{j}_{\mu} = \frac{1}{2\pi}\epsilon^{\mu \nu
\lambda} \partial_{\nu} \tilde{a}_{\lambda}.
\end{align}

In the $U(1) \times U(1) \rtimes Z_2$ Chern-Simons theory, we package
the two gauge fields in the following way: 
\begin{align}
\label{gaugeField} A_\mu = \left(\begin {array}{cc} a_\mu  & 0 \\ 0  &
\tilde{a}_{\mu}\\ \end{array} \right).
\end{align}
The gauge group $G = U(1) \times U(1) \rtimes Z_2$ consists of the
$U(1) \times U(1)$ part, which we can write as 
\begin{align}
U = \left(\begin {array}{cc}
    e^{if}  & 0 \\
    0  & e^{ig}\\
\end{array}
\right),
\end{align}
and the $Z_2$ part, which contains the identity and the non-trivial
element $\sigma_1$: \begin{align} \sigma_1 = \left(\begin {array}{cc}
0  & 1 \\ 1  & 0\\
\end{array}
\right).
\end{align}
Thus, in addition to the usual $U(1) \times U(1)$ gauge symmetry
associated with the two gauge fields, there is a local $Z_2$ gauge
symmetry, which can be thought of in the following way. The space of
physical configurations at a certain space-time point $(x,y,t)$ is to
be described by the unordered pair $(a_{\mu}(x,y,t),
\tilde{a}_{\mu}(x,y,t))$. The action of the $Z_2$ is to interchange
$a_{\mu}(x,y,t)$ and $\tilde{a}_{\mu}(x,y,t)$ at the point $(x,y,t)$.
Physically, we may perhaps envision this as an electron from one layer
and an electron from the other layer being interchanged.  In order to
define a sensible action, we need to be dealing with differentiable
gauge fields. So, we require the gauge fields to be smooth functions
on $M$, thus automatically gauge-fixing the local $Z_2$ and  leaving
behind a residual global $Z_2$ symmetry associated with interchanging
$a$ and $\tilde{a}$ at every point in space-time. In this sense,  we
can use the action given by eqn. $(\ref{lagrangian})$ to describe our
$U(1)\times U(1) \rtimes Z_2$ Chern-Simons theory. 

Although the $U(1)\times U(1)$ Chern-Simons theory and $U(1)\times
U(1) \rtimes Z_2$ Chern-Simons theory formally share the same
Lagrangian, their gauge structure is different.  This is why the same
Lagrangian actually describes two different theories.  This example
demonstrates that the Lagrangian is not a good symbol for a one-to-one
labelling of different topological field theories.

\section{Ground State Degeneracy  for $U(1) \times U(1) \rtimes Z_2$
Chern-Simons Theory}

The first check that a field theory correctly describes a given
topologically ordered phase is whether it correctly reproduces the
ground state degeneracy of the system on surfaces of higher genus.
Accordingly, we begin our study of $U(1) \times U(1) \rtimes Z_2$ by
calculating the ground state degeneracy on a torus. We then calculate
the degeneracy on surfaces of arbitrary genus, from which we deduce
the quantum dimensions of the quasiparticles. Finally, we study the
quasiparticles.

Gauge theory with gauge group $G$ on a manifold $M$ is most generally
defined by starting with a principal $G$ bundle on $M$ and defining
the gauge field, a Lie algebra-valued one-form, as a connection on the
bundle. Often, one is concerned with situations in which $M =
\mathbb{R}^n$, in which case there is a global coordinate system and
the gauge field can be written in coordinates everywhere as $a_\mu
dx^\mu$, where $a_\mu$ is a Lie algebra-valued function on
$\mathbb{R}^n$. In these cases, we do not need to be concerned with
the more general fiber bundle definition in order to compute
quantities of interest. The situation is more complicated in general,
when $M$ does not have a global coordinate system, in which case we
can only locally define $a = a_\mu dx^\mu$ in any given coordinate
chart. In these situations, it is often convenient, when possible, to
view the gauge field as a function defined on $\mathbb{R}^n$, where
$n$ is the dimension of $M$, and to impose suitable periodicity
conditions. This allows us to work in a global coordinate system and
may simplify certain computations. For example, for $U(1)$ gauge
theory on a torus, we can choose to work with a gauge field
$a_\mu(x,y)$ defined over $\mathbb{R}^2$, but with periodic boundary
conditions: 
\begin{equation} 
a_\mu(x,y) = a_\mu(x+L_x,y) = a_\mu(x,y+L_y). 
\end{equation}
In the case where $G = U(1)\times U(1) \rtimes Z_2$, the $Z_2$ gauge
symmetry allows for the possibility of twisted sectors: configurations
in which the gauge field is periodic up to conjugacy by an element of
$Z_2$. On a torus, there are four sectors and the ground state
degeneracy is controlled by the degeneracy within each sector.  In
more mathematical terms, there are four distinct classes of $U(1)
\times U(1) \rtimes Z_2$ bundles on a torus, distinguished by the four
possible elements in the group $(\text{Hom}: \pi_1(T^2) \rightarrow
Z_2)/Z_2$, which is the group of homomorphisms from the fundamental
group of $T^2$ to $Z_2$, mod $Z_2$.  Thus, we can think of
$A_{\mu}(x,y,t)$ as defined on $\mathbb{R}^3$, with the following
periodicity conditions: 
\begin{align} 
A_\mu(x+L_x, y) =
\sigma_1^{\epsilon_x} A_\mu(x,y) \sigma_1^{\epsilon_x} \nonumber \\
A_\mu(x, y + L_y) = \sigma_1^{\epsilon_y} A_\mu(x,y)
\sigma_1^{\epsilon_y} ,
\end{align}
where $\epsilon_x$ and $\epsilon_y$ can each be $0$ (untwisted) or $1$
(twisted).  Furthermore, in each of these sectors, the allowed gauge
transformations $U(x,y)$ take the form (time index is suppressed)
\begin{align}
U(x,y) = \left(\begin {array}{cc}
    e^{if(x,y)}  & 0 \\
    0  & e^{ig(x,y)}\\
\end{array}
\right)
\end{align}
and must preserve the boundary conditions on $A_\mu$:
\begin{align}
U(x+L_x, y) = \sigma_1^{\epsilon_x} U(x,y) \sigma_1^{\epsilon_x}
\nonumber \\
U(x, y + L_y) = \sigma_1^{\epsilon_y} U(x,y) \sigma_1^{\epsilon_y}.
\end{align}
These transform $A_\mu$ in the usual way:
\begin{equation}
A_\mu \rightarrow U A_\mu U^{-1} + i U \partial_\mu U^{-1}.
\end{equation}

The formulation of the theory on higher genus surfaces is similar.  On
a genus $g$ surface, there are $2^{2g}$ different sectors,
characterized by whether there is a $Z_2$ twist along various
non-contractible loops.  Across these twists, the two gauge fields $a$
and $\tilde{a}$ transform into each other.  The gauge transformations
also obey these same twisted boundary conditions; this implies that
the boundary conditions on the gauge fields are preserved under gauge
transformations.

The connection between this formulation and the definition of a
principal $G$-bundle on a compact Riemann surface can be made more
precise by considering local coordinate charts, transition functions,
\it etc \rm, but here we do not pursue any further mathematical
precision. 

\subsection{Ground State Degeneracy on a Torus}

As mentioned above, there are four sectors on a torus, one untwisted
sector and three twisted sectors. We now proceed to compute the ground
state degeneracy in each sector. We follow the approach in
\Ref{WZ9817}, which was applied to continuous and connected gauge
groups. 

\subsubsection{Untwisted Sector}

In the untwisted sector, the ground states are the $Z_2$ invariant
states of a $U(1) \times U(1)$ Chern-Simons theory with the Lagrangian
of eqn. (\ref{lagrangian}). We partially fix the gauge by setting $a_0
= \tilde{a}_0 = 0$. The equations of motion for $a_0$ and
$\tilde{a}_0$, act as constraints that require zero field strength: $f
= \partial_x a_y - \partial_y a_x = 0$ and $\tilde{f} = \partial_x
\tilde{a}_y - \partial_y \tilde{a}_x = 0$.  This implies that
gauge-inequivalent configurations are completely specified by the
holonomies of the gauge fields around non-contractible loops of the
torus, $\oint a \cdot dl$ and $\oint \tilde{a} \cdot dl$. This is a
special case of the more general statement that flat $G$-bundles are
characterized by $(\text{Hom}: \pi_1(M) \rightarrow G)/G$. We can
parameterize this configuration space in the following way.
\begin{align}
a_1(x_,y_,t) = \frac{2 \pi}{L} X(t) & \ \ \ \ \tilde{a}_1(x_,y_,t) = \frac{2 \pi}{L} \tilde{X}(t)
\nonumber\\
a_2(x_,y_,t) = \frac{2 \pi}{L} Y(t) & \ \ \ \ \tilde{a}_2(x_,y_,t) = \frac{2 \pi}{L} \tilde{Y}(t) 
\end{align}
The large gauge transformations $a \rightarrow a + i U^{-1} \partial
U$ with $U(x, y) = e^{2\pi i m x/L + 2\pi i n y/L}$ take $(X,Y)
\rightarrow (X+m, Y+n)$.  Thus $(X,Y)$ and $(\tilde{X},\tilde{Y})$
take values on a torus.  Substitution into the action yields, up to
total time derivatives, 
\begin{equation} 
L = 2 \pi k (X \dot{Y} +
\tilde{X} \dot{\tilde{Y}}) + 2 \pi (k-l) (\tilde{X} \dot{Y} + X
\dot{\tilde{Y}}) .
\end{equation}
The Hamiltonian vanishes. The momenta conjugate to $Y$ and $\tilde{Y}$ are
\begin{align}
p_Y = \frac{\delta L}{\delta \dot{Y}} = 2 \pi k X + 2 \pi (k-l) \tilde{X},
\nonumber \\
p_{\tilde{Y}} = \frac{\delta L}{\delta \dot{\tilde{Y}}} = 2 \pi k \tilde{X} + 2 \pi (k-l) X .
\end{align}
The wave functions for this system can be written as a sum of plane waves:
\begin{equation}
\psi(Y,\tilde{Y}) = \sum_{n,m} c_{n,m} e^{i 2 \pi n Y + i 2 \pi m \tilde{Y}} .
\end{equation}
In momentum space, the wavefunction becomes
\begin{equation}
\phi(p_Y, p_{\tilde{Y}}) = 
\sum_{n,m} c_{n,m} \delta (p_Y - 2 \pi n) \delta (p_{\tilde{Y}} - 2 \pi m),
\end{equation}
or, equivalently,
\begin{equation}
\varphi(X,\tilde{X}) = \sum_{n,m} c_{n,m} 
\delta(k X + (k-l) \tilde{X} - n) \delta (k \tilde{X} + (k-l) X - m).
\end{equation}
Using the fact that $X \sim X + 1$ and $\tilde{X} \sim \tilde{X} + 1$,
we find that
\begin{equation}
\label{id1}
c_{n,m} = c_{n-k,m-k+l} = c_{n-k+l,m-k}.
\end{equation}
There are $l(2k-l)$ independent coefficients
$c_{n,m}$, which explains why the $(k, k, k-l)$ quantum Hall state has
a degeneracy of $l(2k-l)$ on a torus.

We can label the quantum states by $| n, m \rangle$. The ground states
in our $U(1)\times U(1)\rtimes Z_2$ theory will be the $Z_2$ invariant
subspace of this Hilbert space; it will contain the diagonal states
$|n, n \rangle$ and ones of the form $|n, m \rangle + |m, n \rangle$.
A simple count of the $Z_2$ invariant states, using the identifications
(\ref{id1}) yields a total of
\begin{equation}
(l+1)(2k-l)/2
\end{equation}
states in this untwisted sector. 

\subsubsection{Twisted Sectors}

There are three $Z_2$ twisted sectors, corresponding to twisting in either the 
$x$ direction, the $y$ direction, or both. Since modular transformations, \it i.e. \rm
diffeomorphisms that are not continuously connected to the identity, 
are symmetries that can take one twisted sector to another, we expect 
that all twisted sectors should have the same degeneracy. This can be 
verified explicitly by computing the degeneracy in each case. Here 
we will only consider the case where the gauge fields are twisted in 
the $y$ direction. More precisely this means that the gauge fields obey
the following boundary conditions:
\begin{align}
a_i(x,y+L) = \tilde{a}_i(x,y) \ \ \ \ \ \tilde{a}_i(x,y+L) = a_i(x,y) 
\nonumber \\
a_i(x+L,y) = a_i(x,y) \ \ \ \ \ \tilde{a}_i(x+L,y) = \tilde{a}_i(x,y) 
\end{align}

Given these twisted boundary conditions, we can consider a new field 
$c_{\mu}(x,y)$ defined on a space that is doubled in length
in the $y$ direction:
\begin{equation}
c_{\mu}(x,y) = \left\{ 
  \begin{array}{ll}
    a_{\mu}(x,y) &  0 \leq y \leq L \\
    \tilde{a}_{\mu}(x,y-L) &  L \leq y \leq 2L
  \end{array} \right.
\end{equation}
Observe that $c$ has the periodicity
\begin{equation}
\label{cPeriodicity}
c_{\mu}(x,y) = c_{\mu}(x+L,y) = c_{\mu}(x,y+2L).
\end{equation}
The allowed gauge transformations that act on $c_i$ are
of the form $W(x,y) = e^{ih(x,y)}$, where
$W(x,y)$ need only be periodic on the doubled torus:
\begin{align}
W(x+L,y) = W(x,y+2L) = W(x,y)
\end{align}
$c$ transforms as a typical $U(1)$ gauge field:
\begin{equation}
c\rightarrow c - \partial h.
\end{equation}
In particular, there are large gauge transformations 
$W(x,y) = e^{i\frac{2\pi m}{L} x + i \frac{2 \pi n}{2L} y}$
that change the zero-mode of $c_i$:
\begin{equation}
c_i \rightarrow c_i + \frac{2\pi m}{L} + \frac{2 \pi n}{2L}
\end{equation}

In terms of $c$, the Lagrangian becomes
\begin{equation}
L = \int_0^L dx \int_0^{2L} dy (\frac{k}{4\pi} c \partial c + \frac{k-l}{4\pi} c(x,y) \partial c(x,y-L) )
\end{equation}
Note that this lagrangian is actually non-local in the field $c$, but
this does not pose any additional difficulty. We can set temporal
gauge $c_0 = 0$, \it i.e. \rm $a_0 = \tilde{a}_0 = 0$, and view the
equation of motion for $c_0$ as a constraint that forces the field
strength for $c$ to be zero. Thus, the gauge-inequivalent
configurations can be parameterized as
\begin{align}
c_i(x,y,t) = \frac{2 \pi}{L_i} X_i(t),
\end{align}
where $L_1 = L$ and $L_2 = 2 L$. Inserting this expansion into the
Lagrangian gives, up to total time derivatives,
\begin{align}
L = 2 \pi (2k-l) X_1 \dot{X}_2 .
\end{align}
Due to the existence of the large gauge transformations, we find that
the zero-modes $X_i$ take values on a torus:
\begin{equation}
(X_1,X_2) \sim (X_1 + 1, X_2) \sim (X_1, X_2 + 1).
\end{equation}
Thus, using the same techniques used in the previous section, we
conclude that the ground state degeneracy in this sector is $2k - l$.
There are three different twisted sectors, so we find in
total 
\begin{equation}
3 (2k - l)
\end{equation}
states in the twisted sectors of the $U(1)\times U(1)\rtimes Z_2$
theory.

\subsubsection{Total Ground State Degeneracy on Torus}

Adding the degeneracies from the twisted and the untwisted
sectors, we find that the total ground state degeneracy on a torus in
$U(1)\times U(1)\rtimes Z_2$ theory is
\begin{equation}
\label{torusDeg}
\text{ Ground State Deg. on Torus} = (l+7)(2k - l)/2 .
\end{equation}

For $l=2$, the filling fraction is $\nu = \frac{1}{k-1}$ and the above
formula gives $9(k-1)$ states on a torus. Compare this to the torus
degeneracy of the $\nu = \frac{1}{k-1}$ Pfaffian state, which is $3
(k-1)$. We see that the $U(1) \times U(1) \rtimes Z_2$ Chern-Simons
theory for $l=2$ has a torus ground state degeneracy that is three
times that of the Pfaffian state.  So the  $U(1) \times U(1) \rtimes
Z_2$ Chern-Simons theory for $l=2$ cannot directly describe the
Pfaffian state. In Appendix \ref{torusDiscussion}, we argue that, for
$l=2$,  $U(1) \times U(1) \rtimes Z_2$ Chern-Simons theory describes
the Pfaffian state plus an extra copy of the Ising model.

For $l=3$, the filling fraction is $\nu = \frac{2}{2k-3}$ and
(\ref{torusDeg}) gives $5(2k-3)$ ground states on a torus.  The $\nu =
\frac{2}{2k-3}$ $Z_4$ parafermion state also gives rise to same torus
degeneracy of $5(2k-3)$.  Thus, we would like to propose that the
$U(1) \times U(1) \rtimes Z_2$ Chern-Simons theory for $l=3$ describes
the $Z_4$ parafermion quantum Hall states.  As a more non-trivial
check on these results, we now turn to the calculation of the ground
state degeneracy on surfaces of arbitrary genus. 

\subsection{Ground State Degeneracy for genus $g$}

The ground state degeneracy on a genus $g$ surface of the $Z_4$
parafermion quantum Hall state at filling fraction $\nu =
\frac{2}{2k-3}$ is given by\cite{BW0932} 
\begin{equation}
\label{Z4deg}
(k-3/2)^g 2^{g-1}[ (3^g + 1) + (2^{2g}-1) (3^{g-1} + 1)] .
\end{equation}
Note that the second factor, $2^{g-1}[ (3^g + 1) + (2^{2g}-1) (3^{g-1}
+ 1)]$, is the dimension of the space of conformal blocks on a genus
$g$ surface in the $Z_4$ parafermion CFT (see (\ref{dimVg})). 
The degeneracy for the corresponding quantum Hall state is 
$(k-3/2)^g = \nu^{-g}$ times this factor.

\begin{figure}[tb]
\centerline{
\includegraphics[scale = 0.9]{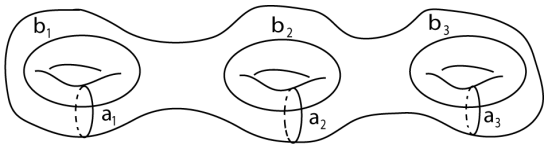}
}
\caption{\label{homologycycles}
Canonical homology basis for $\Sigma_g$. }
\end{figure}

Let us consider the ground state degeneracy on a genus $g$ surface for the
$U(1) \times U(1) \rtimes Z_2$ Chern-Simons theory. Let $\{a_i\}$ and 
$\{b_i\}$, with $i = 1, \cdots, g$ be a basis for the homology cycles 
(see Figure \ref{homologycycles}). The $a_i$ ($b_i$) do not 
intersect each other, while $a_i$ and $b_j$ intersect if $i = j$. That is,
the $a_i$ and $b_i$ form a canonical homology basis. There can be 
a $Z_2$ twist along any combination of these non-contractible loops. Thus 
there are $2^{2g}$ different sectors; one of them is untwisted while the 
other $2^{2g} -1$ sectors are twisted. Let us first analyze the untwisted sector. 

It is known that the $(k, k, k-l)$ bilayer FQH states, which are described by 
the $U(1) \times U(1)$ Chern-Simons theory of eqn. (\ref{lagrangian}) have a 
degeneracy of $(\text{det}{K})^g$, where the $K$-matrix is:
\begin{align}
K = \left(\begin {array}{cc}
    k  & k-l \\
    k-l  & k\\
\end{array}
\right).
\end{align}
Thus the degeneracy for these bilayer states is $l^g(2k-l)^g$. These states may be
written as:
\begin{equation}
\otimes_i |n_i, m_i \rangle,
\end{equation}
where the $n_i$ and $m_i$ are integers,
$i = 1, \cdots, g$, and with the identifications (see (\ref{id1}))
\begin{equation}
(n_i,m_i) \sim (n_i + k - l, m_i +k) \sim (n_i +k,m_i +k-l)
\end{equation}
for each $i$. The action of the $Z_2$ on these states is to take 
\begin{equation}
\otimes_i |n_i, m_i \rangle \rightarrow \otimes_i |m_i, n_i \rangle.
\end{equation}
We must project onto the $Z_2$ invariant states. There are $(2k-l)^g$ diagonal
states of the form $\otimes_i |n_i, n_i \rangle$. These are invariant under the $Z_2$.
There are $l^g(2k-l)^g - (2k-l)^g$ off-diagonal states, and exactly half of them are
$Z_2$ invariant. This gives a total of 
\begin{equation}
(l^g + 1) (2k - l)^g / 2 = (l^g + 1) (k - l/2)^g 2^{g-1}
\end{equation}
different states, which for $l = 3$ corresponds to the first term of (\ref{Z4deg}).

\begin{figure}[tb]
\centerline{
\includegraphics[scale=0.9]{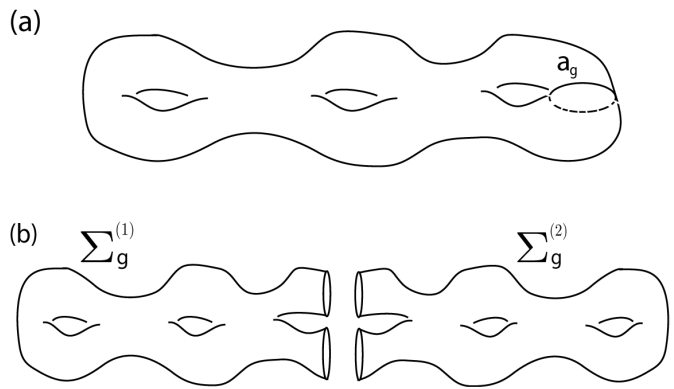}
}
\caption{
\label{doublecover}
(a) A single twist along the $a_g$ direction. 
(b) Take two copies of $\Sigma_g$, cut them along the $a_g$ cycle, and glue them together as shown.
This yields a genus $2g-1$ surface. For the case $g=3$, we see explicitly that a genus $5$ surface is obtained. }
\end{figure}

Now consider the twisted sectors. To begin, suppose that there is a $Z_2$ twist along
the $a_g$ cycle, and no twists along any of the other cycles. Let $\Sigma_g$ refer to
the genus $g$ surface. Let us consider the double cover $\hat \Sigma_{2g-1}$ of $\Sigma_g$,
which is a genus $2g-1$ surface. It can be constructed as follows. Take two copies
of $\Sigma_g$, referred to as $\Sigma_g^1$ and $\Sigma_g^2$, and cut both of them 
along their $a_g$ cycle. Gluing them together in such a way that each 
end of the cut on one copy lands on the opposite end of the cut on the other copy
leaves the $2g-1$ surface $\hat \Sigma_{2g-1}$ (see Figure \ref{doublecover}). 
The sheet exchange $R$ is a map from $\hat \Sigma_{2g-1}$ to itself 
that satisfies $R \circ R = 1$ and which takes $\Sigma_g^1 \rightarrow \Sigma_g^2$ and 
vice versa.\cite{AS9207} We can now define a new, continuous gauge field $c$ on $\hat \Sigma_{2g-1}$ as follows:
\begin{equation}
\label{cdef}
c(p) = \left\{ 
  \begin{array}{ll}
    a(p) &  p \in \Sigma_g^1 \\
    \tilde{a}(R(p)) &  p \in \Sigma_g^2
  \end{array} \right.
\end{equation}
Notice that because the gauge transformations get twisted also, $c$ now behaves
exactly as a typical $U(1)$ gauge field on a genus $2g-1$ surface. In particular,
there are large gauge transformations which change the value of $\oint_{\alpha_i} c \cdot dl$
or $\oint_{\beta_i} c \cdot dl$ by $2\pi$. 

In terms of $c$, the action (\ref{lagrangian}) becomes
\begin{equation}
\label{LagrangianG}
L = \int_{\hat\Sigma_{2g-1}} \frac{k}{4\pi} c(p) \partial c(p) + \frac{k-l}{4 \pi} c(p) \partial c(R(p)).
\end{equation}
In terms of $c$, the Lagrangian is non-local, however this poses no difficulty. Fixing the
gauge $c_0 = 0$, the equation of motion for $c_0$ is a constraint that enforces $c$ to have
zero field strength; that is, $c$ is a flat connection. 

Let $\{ \alpha_i\}$ and $\{\beta_i\}$ be a basis of canonical homology cycles on $\hat \Sigma_{2g-1}$, with
$i = 1, \cdots, 2g-1$. We can choose $\alpha_i$ and $\beta_i$ in such a way that the sheet exchange $R$ acts on these 
cycles as follows:
\begin{align}
\label{sheetExchangeAction}
R \alpha_i = \alpha_{i + g - 1 }, \;\;\;\; R \beta_i = \beta_{i + g - 1}
\nonumber \\
R \alpha_{2g - 1} = \alpha_{2g-1}, \;\;\;\; R \beta_{2g-1} = \beta_{2g-1},
\end{align}
where $i = 1, \cdots, g-1$. The dual basis is the
set of one-forms $\omega_i$ and $\eta_i$, which satisfy
\begin{align}
\int_{\alpha_i} \omega_j = \delta_{ij} \;\;\;\; \int_{\beta_i} \omega_j = 0,
\nonumber \\
\int_{\alpha_i} \eta_j = 0 \;\;\;\; \int_{\beta_i} \eta_j = \delta_{ij}.
\end{align}
Since $c$ must be a flat connection, we can parametrize it as
\begin{equation}
c = c_1 dx^1 + c_2 dx^2 = 2\pi (x^i \omega_i + y^i \eta_i).
\end{equation}
Two connections $c$ and $c'$ are gauge-equivalent if 
\begin{equation}
x'^i - x^i = \text{ integer }, \;\;\;\; y'^i - y^i = \text{ integer }.
\end{equation}
Furthermore, from the definition of $c$ (eqn. \ref{cdef}), we see that
the $Z_2$ action is the same as the action of the sheet exchange $R$: 
\begin{equation}
(x^i,y^i) \rightarrow (x^{R(i)}, y^{R(i)}),
\end{equation}
where 
\begin{equation}
\label{Rdef}
R(i) = \left\{ 
  \begin{array}{lll}
    i+g-1 &  \text{for } & i = 1, \cdots, g-1 \\
    i-g+1 &  \text{for } & i = g, \cdots, 2g-2 \\
    2g-1 & \text{for } & i = 2g-1
  \end{array} \right.
\end{equation}
Substituting into the action (\ref{LagrangianG}) and using the fact that
$\int_{\hat \Sigma_{2g-1}} \omega_j \wedge \eta_k = \delta_{jk}$ and
$\int_{\hat \Sigma_{2g-1}} \omega_j \wedge \omega_k = \int_{\hat \Sigma_{2g-1}} \eta_j \wedge \eta_k = 0$,
we obtain
\begin{equation}
L = 2 \pi k y^i \dot{x}^i + 2\pi (k-l)  y^i \dot{x}^{R(i)}.
\end{equation}
Apart from the variables with $i = 2g-1$, this action looks like
the action for a bilayer $(k,k,k-l)$ state on a genus $g-1$ surface.
Therefore, we can easily deduce that quantizing this system before 
imposing the invariance under the $Z_2$ action gives 
$l^{g-1}(2k-l)^{g-1} \times (2k-l)$ different states. The extra factor $2k-l$ 
comes from the variables with $i = 2g-1$, which independently behave as the zero-modes
of a $U(1)_{2k-l}$ C.S. theory on a torus. We can write the states as
\begin{equation}
|n_{2g-1} \rangle \otimes_i |n_i, n_{R(i)} \rangle,
\end{equation} 
for $i = 1, \cdots, g-1$ and with the identifications 
\begin{equation}
n_{2g-1} \sim n_{2g-1} + 2k-l,
\end{equation}
\begin{align}
(n_i,n_{R(i)}) &\sim (n_i + k,n_{R(i)}+k-l) 
\nonumber \\
& \sim (n_i + k - l, n_{R(i)} + k).
\end{align}
Note the $n_i$ are all integer. Now we must project onto the 
$Z_2$ invariant sector. The action of the $Z_2$ is to take
\begin{equation}
|n_{2g-1} \rangle \otimes_i |n_i, n_{R(i)} \rangle \rightarrow |n_{2g-1} \rangle \otimes_i |n_{R(i)}, n_i \rangle.
\end{equation}
Suppose $n_i = n_{R(i)}$ for each $i$. Such states are already $Z_2$ invariant; there are
$(2k-l) \times (2k-l)^{g-1}$ of them. The remaining states for which $n_i \neq n_{R(i)}$ for at least
one $i$ always change under the $Z_2$ action. The $Z_2$ invariant combination is
\begin{equation}
|n_{2g-1} \rangle \otimes_i (|n_i, n_{R(i)} \rangle + |n_{R(i)}, n_i \rangle).
\end{equation}
There are $(2k-l) \times \frac{l^{g-1} (2k-l)^{g-1} - (2k-l)^{g-1}}{2}$ of these. In total therefore,
there are
\begin{equation}
\label{twistedSectorG}
(2k-l)^g \frac{l^{g-1} + 1}{2} = (k - l/2)^g (l^{g-1} + 1) 2^{g-1}
\end{equation}\
states in this particular twisted sector.

\begin{figure}[tb]
\centerline{
\includegraphics[scale = 0.9]{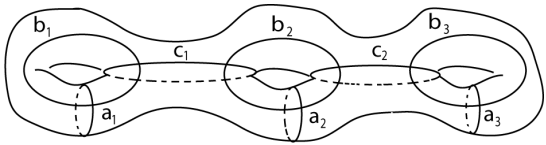}
}
\caption{\label{dehntwists}
Canonical homology basis for $\Sigma_g$. }
\end{figure}

Now it turns out that each of the $2^{2g} - 1$ twisted
sectors (which generically has many $Z_2$ twists along many
different non-contractible loops) yield the same number of 
ground states as the sector in which there is
a single twist along just the $a_g$ cycle. 
One can understand this by considering the modular 
group, or mapping class group, of $\Sigma_g$. 
This is the group of diffeomorphisms on $\Sigma_g$
modulo those that are continuously connected to the identity.
They are generated by ``Dehn twists,'' which correspond
to cutting the surface along some non-contractible loop,
rotating one side by $2 \pi$, and gluing the two sides back together.
The mapping class group of $\Sigma_g$ can be generated
by Dehn twists along the loops $a_i$, $b_i$, and $c_i$,
shown in Figure \ref{dehntwists}. Elements of the mapping class group
are symmetries of the topological field theory,
which means that they are represented by unitary operators 
on the quantum Hilbert space. In particular, the dimension
of the space of states for a given twisted sector is
equivalent to that of a different twisted sector if they 
can be related by the action of an element of the mapping
class group. 

In the following we sketch how, using Dehn twists,
one can go from any arbitrary twisted sector to the sector in which
there is a single $Z_2$ twist along only the $a_g$ cycle. 

First note that a $Z_2$ twist along some cycle $\gamma$ 
is equivalent to having a $Z_2$ twist along $-\gamma$,
and that a $Z_2$ twist along $\gamma + \gamma$ is equivalent
to having no $Z_2$ twist at all. Since we are here concerned only
with the properties of the $Z_2$ twists, we use these 
properties in the algebra below. In other words,
the algebra below will be defined over $Z_2$
because we are only concerned with $Z_2$ twists along various cycles.

Let us call $A_i$, $B_i$, and $C_i$ the Dehn twists that act 
along the $a_i$, $b_i$, and $c_i$ cycles.
Notice that a $Z_2$ twist along $a_i$ and $a_{i+1}$
is equivalent to a $Z_2$ twist along $c_i$. Let us
consider the action of $A_i$, $B_i$, and $C_i$ on
$Z_2$ twists along the $a_i$ and $b_i$ cycles. 
\begin{align}
A_i:\; &a_i \rightarrow a_i
\nonumber \\
 &b_i\rightarrow a_i + b_i,
\nonumber \\
B_i:\; &a_i \rightarrow a_i + b_i 
\nonumber \\
& b_i \rightarrow b_i.
\nonumber \\
C_i:\; &a_i \rightarrow a_i 
\nonumber \\
&  b_i \rightarrow b_i + c_i = b_i + a_i + a_{i+1}
\nonumber \\
&a_{i+1} \rightarrow a_{i+1}  
\nonumber \\
& b_{i+1} \rightarrow b_{i+1} + c_i = b_{i+1} + a_i + a_{i+1}
\end{align}
$Z_2$ twists along all other cycles are left unchanged. 
Notice in particular that $A_i^{-1}B_i: a_i \rightarrow b_i$,
so that a $Z_2$ twist along $a_i$ is equivalent to one
along $a_i + b_i$, which is also equivalent to one along
$b_i$. As a result, we can see that the
configuration of $Z_2$ twists can be labelled only by considering
which of the $g$ handles have any twists at all. Furthermore, since
we can rearrange the holes without changing the topology, the
configuration of $Z_2$ twists is actually labelled by considering
how many of the $g$ handles have twists. 

Suppose that two of the $g$ handles have $Z_2$ twists. Since we have
freedom to rearrange the holes, we can consider the situation in which
two neighboring handles each have a $Z_2$ twist. Since twists
along $a_i$, $a_i + b_i$, and $b_i$ are all
equivalent, let us suppose that one handle has a twist along its
$b$ cycle, while the other handle has a twist along its $a$
cycle. That is, we are considering the situation in which there is a
twist along $b_i + a_{i+1}$. Now, performing the Dehn 
twist $C_i$, we have:
\begin{align}
C_i : b_i + a_{i+1} &\rightarrow b_i + a_i + a_{i+1} + a_{i+1} 
\nonumber \\
&= b_i + a_i.
\end{align}
Thus we see that the case with $Z_2$ twists for two handles is equivalent to that
for a $Z_2$ twist along a single handle. From this, it follows that the case with $n$
handles having $Z_2$ twists is equivalent to the case where only a single handle
has a $Z_2$ twist. 

Therefore, under actions of the Dehn twists, any arbitrary 
twisted sector goes into the sector in which there is a single 
twist along the $a_g$ cycle. This means that the dimension of the
Hilbert space is the same for each of the $(2^{2g} - 1)$ twisted sectors,
and in particular is equal to that for the sector in which there is a single
twist along $a_g$. We computed that situation explicitly (see eqn. \ref{twistedSectorG}),
so we can conclude that the number of ground states on a genus $g$ 
surface for the $U(1) \times U(1) \rtimes Z_2$ Chern-Simons theory is:
\begin{equation}
S_g(k,l) =  (k - l/2)^g 2^{g-1} [ (l^g + 1)   + (2^{2g} - 1)  (l^{g-1} + 1) ].
\end{equation}
For $l = 3$, this corresponds to the degeneracy of the $Z_4$ parafermion
quantum Hall state that we expect from a CFT calculation (see eqn. \ref{Z4deg}). When $l=2$, we get
\begin{equation}
S_g(k,2) = (k-1)^g [2^{g-1} (2^g + 1)]^2,
\end{equation}
which corresponds to the degeneracy of the $\nu = \frac{1}{k-1}$ Pfaffian
quantum Hall state times an extra factor of $2^{g-1} (2^g + 1)$, which is the
dimension of the space of conformal blocks of the Ising CFT 
on a genus $g$ surface. This again confirms the notion that for $l=2$, 
this theory corresponds to the Pfaffian state with an extra copy of the Ising model. 

\section{Quantum Dimensions of Quasiparticles from Ground State Degeneracy}

In the last section we found the ground state degeneracy, $S_g$, of the $U(1) \times U(1) \rtimes Z_2$
Chern-Simons theory on a surface of genus $g$. From $S_g$ we can deduce some topological properties
of the quasiparticles. It is well known for example that $S_1$, the ground state degeneracy on a torus, 
is equal to the number of topologically distinct quasiparticles. Here we show that from $S_g$
we can also obtain the quantum dimensions of each of the quasiparticles. 

The quantum dimension $d_\gamma$ of a quasiparticle denoted by $\gamma$ has the following meaning.
For $n$ quasiparticles of type $\gamma$ at fixed positions, the dimension of the Hilbert space 
grows as $d_\gamma^n$. For Abelian quasiparticles at fixed positions, there is no degeneracy of states,
so the quantum dimension of an Abelian quasiparticle is one. The quantum dimension $d_\gamma$ can
be obtained from the fusion rules of the quasiparticles, $N_{\gamma \gamma'}^{\gamma''}$: $d_\gamma$
is the largest eigenvalue of the fusion matrix $N_{\gamma}$, where
$(N_{\gamma})^{\gamma''}_{\gamma'} = N_{\gamma \gamma'}^{\gamma''}$. From the quantum dimensions $d_{\gamma}$, 
we can obtain $S_g$ through the formula\cite{V8860,BW0932} 
\begin{equation}
S_g = D^{2(g-1)} \sum_{\ga=0}^{N-1} d_\ga^{-2(g-1)},
\end{equation}
where $N$ is the number of quasiparticles,  $d_\ga$ is the quantum 
dimension of quasiparticle $\ga$ and $D = \sqrt{\sum_\ga d_\ga^2}$
is the ``total quantum dimension.'' Remarkably, this formula also implies that if we
know $S_g$ for any $g$, then we can uniquely determine all of the quantum dimensions
$d_\ga$. To see how, let us first order the quasiparticles so that $d_{\ga+1} \geq d_{\ga}$.
Notice that the identity has unit quantum dimension: $d_0 = 1$, and suppose that 
$d_i = 1$ for $i = 0, \cdots, i_0$ ($i_0 \geq 0$), $d_{i_0 + 1} > 1$. Now consider
\begin{align}
\lim_{g \rightarrow \infty} \frac{S_{g+1}}{S_g} &= D^2 \lim_{g \rightarrow \infty} 
\frac{i_0 + \sum_{\ga = i_0 +1}^{N-1}d_{\ga}^{-2g}}{i_0 + \sum_{\ga = i_0 +1}^{N-1}d_{\ga}^{-2(g-1)}} 
\nonumber \\
&= D^2.
\end{align}
We see that the total quantum dimension $D$ can be found by computing 
$\lim_{g \rightarrow \infty} \frac{S_{g+1}}{S_g}$. Now define
\begin{equation}
\tilde S_g^{(1)} \equiv \frac{S_g}{D^{2(g-1)}} - 1 = \sum_{\ga = 1}^{N-1} d_{\ga}^{-2(g-1)},
\end{equation}
and suppose that $d_{1}, \cdots, d_{i_1}$ all have the same quantum dimension. Now consider the following limit. 
\begin{align}
\lim_{g \rightarrow \infty} \frac{\tilde S_{g+1}^{(1)}}{\tilde S_g^{(1)}} &= 
\lim_{g \rightarrow \infty} \frac{d_1^{-2g}(i_1 + \sum_{\ga = i_1 + 1}^{N-1} d_{\ga}^{-2g})}{d_1^{-2(g-1)}(i_1 + \sum_{\ga = i_1 + 1}^{N-1} d_{\ga}^{-2(g-1)})}
\nonumber \\
&= d_1^{-2}. 
\end{align}
We see that $d_1$ can be determined by computing $\lim_{g \rightarrow \infty} \frac{\tilde S_{g+1}^{(1)}}{\tilde S_g^{(1)}}$. 
This allows one to define
\begin{equation}
\tilde S_g^{(2)} \equiv \tilde S_g^{(1)} - d_1^{-2(g-1)} = \sum_{\ga = 2}^{N-1} d_\ga^{-2(g-1)},
\end{equation}
and in turn we find $d_2^{-2} = \lim_{g \rightarrow \infty} \frac{\tilde S_{g+1}^{(2)}}{\tilde S_g^{(2)}}$.
Proceeding in this way, one can obtain $d_i$, then define 
\begin{equation}
\tilde S_g^{(i+1)} \equiv \tilde S_g^{(i)} - d_i^{-2(g-1)} = \sum_{\ga = i+1}^{N-1} d_\ga^{-2(g-1)},
\end{equation}
and then compute $d_{i+1}$ from $\tilde S_g^{(i+1)}$:
\begin{equation}  
d_{i+1}^{-2} = \lim_{g \rightarrow \infty} \frac{\tilde S_{g+1}^{(i+1)}}{\tilde S_g^{(i+1)}}.
\end{equation}
Thus we can see that in this way all of the quantum dimensions of the quasiparticles can be obtained from the formula
for the ground state degeneracy on a genus $g$ surface. 

Carrying out this procedure for the $U(1) \times U(1) \rtimes Z_2$ Chern-Simons 
theory, we find that when $l < 4$, the quantum dimensions of the quasiparticles take
one of three different values. $2(2k-l)$ of them have quantum dimension 1, $2(2k-l)$
of them have quantum dimension $\sqrt{l}$, and the remaining $(l-1)(2k-l)/2$
of them have quantum dimension $2$. The total quantum dimension is
\begin{align}
D^2 = 4l(2k-l).
\end{align}
For $l=3$ this coincides exactly with the quantum
dimensions of the quasiparticles in the $\nu = \frac{2}{2k-3}$ $Z_4$ parafermion FQH
states.

\section{Quasiparticles}

When we refer to quasiparticles in a Chern-Simons theory, we are referring to
topological defects in the configuration of the gauge fields. For instance, 
for a Chern-Simons theory at level $k$ with a simple Lie group $G$, a quasiparticle is represented by
a unit of flux in an integrable representation of the affine Lie algebra $\hat{g}_k$, 
where $g$ is the Lie algebra of $G$. The partition function of the Chern-Simons theory in the presence
of external sources of quasiparticles is
\begin{equation}
Z( \{C_i, R_i\}) = \int \mathcal{D}A \prod_i W_{R_i}(C_i) e^{i S_{c.s.}[A]},
\end{equation}
where the Wilson loop operator $W_R(C)$ is defined as
\begin{equation}
W_R(C) = Tr_{R} \mathcal{P} e^{i \oint_C A \cdot dl}.
\end{equation}
$Tr_{R}$ is a trace in the representaton $R$, $\mathcal{P}$ refers to path-ordering, 
and $C$ is a loop describing the world-line of the quasiparticle. Furthermore, the action
of the quantum operator $\hat{W}_{R_i}(C)$ is to take one ground state to another when
$C$ is a non-contractible loop in space. 

In the $U(1) \times U(1) \rtimes Z_2$ Chern-Simons theory, there are several types of
quasiparticles to consider. Some of the quasiparticles are related to the Wilson
loop operators for the $U(1)$ gauge fields; some are neutral under the $Z_2$ gauge
field while others carry $Z_2$ charge.  There are also $Z_2$ vortices, which we explicitly
analyze in the following section.

\subsection{$Z_2$ Vortices}

One basic excitation in a theory with a $Z_2$ gauge symmetry is a $Z_2$ vortex. In the context
of  $U(1)\times U(1) \rtimes Z_2$ Chern-Simons theory, a $Z_2$ vortex is, roughly speaking,
a point around which the $U(1)$ gauge fields transform into each other. Here we compute 
the degeneracy of states in the presence of $n$ pairs of $Z_2$ vortices at fixed positions; 
we find that this degeneracy grows like $l^{n}$, and therefore the $Z_2$ vortices can be
identified with the non-Abelian quasiparticles with quantum dimension $\sqrt{l}$. We can 
in fact obtain the formula for the degeneracy more precisely and find that it agrees
exactly, for $l = 3$, with results from the $Z_4$ parafermion FQH states. 

The basic idea is that a sphere with $n$ pairs of $Z_2$ vortices can be related to a
$U(1)_l$ Chern-Simons theory on a genus $g = n-1$ Riemann surface. We will find that 
the $Z_2$ invariant subspace of this theory has $(l^{n-1} + 1)/2$ states
while the $Z_2$ non-invariant subspace has $(l^{n-1} - 1)/2$ states when $l$ is odd. 

We may define a pair of $Z_2$ vortices more precisely as a one-dimensional closed sub-manifold $\gamma$ of 
our spatial 2-manifold $M_0$. The two boundary points of $\gamma$ are thought of as the 
location of the $Z_2$ vortices. The gauge field $A_\mu$ is defined on $M = M_0 \backslash \gamma$, with 
the following boundary conditions along $\gamma$:
\begin{equation}
\lim_{p \rightarrow p^{\pm}_0} A_\mu (p) = \lim_{p \rightarrow p^{\mp}_0} \sigma_1 A_\mu(p) \sigma_1
\end{equation}
for every point $p_0 \in \gamma$. The limit $p \rightarrow p_0^{+ (-)}$ means that 
the limit is taken approaching one particular side (or the other) of $\gamma$.

\begin{figure}[tb]
\centerline{
\includegraphics[scale=0.5]{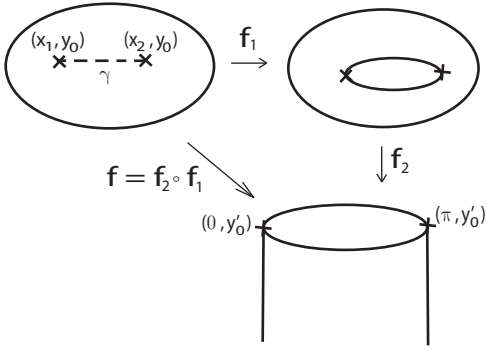}
}
\caption{\label{diffeoFigureBC}
Consider the diffeomorphism $f$, which takes the  neighborhood of a pair of $Z_2$ vortices
to the end of a cylinder. We can imagine $f$ as a composition of two maps, the first which expands the 
cut $\gamma$ to a hole, and a second one which maps the result to the end of a cylinder. }
\end{figure}

Consider the action of a diffeomorphism $f: M \rightarrow M$, which takes
$p \rightarrow p' = f(p)$. The Chern-Simons action is a topological 
invariant and is therefore invariant under diffeomorphisms. 
However, the gauge fields transform along with the coordinates,
which means that the boundary conditions at the boundary of $M = M_0 \backslash \gamma$ 
will change. Let us determine how the boundary conditions on $A$ change under the
action of the diffeomorphism $f$, which acts in the way indicated in Figure \ref{diffeoFigureBC}
in the neighborhood of a pair of $Z_2$ vortices connected by $\gamma$.

Choosing a coordinate chart in the neighborhood of a pair of $Z_2$ vortices, we can write the 
action of $f$ as:
\begin{align}
\label{coordinateChange}
 x^\mu &\rightarrow x'^\mu,
\nonumber \\
 a_{\mu} &\rightarrow a'_\mu = \frac{\partial x^\nu}{\partial x'^\mu} a_\nu.
\end{align}
Let us choose the coordinates $x^\mu$ such that (see Figure \ref{diffeoFigureBC})
\begin{equation}
\gamma = \{(x,y_0) | x_1 \leq x \leq x_2\}.
\end{equation}
The two $Z_2$ vortices are located at the two ends of $\gamma$ and
$f$ maps the neighborhood of these $Z_2$ vortices to the end of a cylinder;
the boundary $M$ in this neighborhood gets mapped to a circle. In terms of the 
new coordinates $x'^\mu$, this neighborhood of $M$ gets mapped to
\begin{equation}
\{(x',y') | y' < y_0', x' \in \mathbb{R} \: \% \: 2\pi \}. 
\end{equation}
The location of the $Z_2$ vortices in the new coordinates 
is taken to be at $(0, y'_0)$ and $(\pi, y_0')$. Fix some 
small $\epsilon > 0$. Let us choose an $f$ that takes 
\begin{equation}
(x_0, y_0 \pm \epsilon) \rightarrow (\mp x_0', y_0'-\epsilon)
\end{equation}
for $x_1 < x_0 < x_2$. It is easy to see that as $\epsilon$ is taken to zero, we have:
\begin{equation}
\lim_{\epsilon \rightarrow 0^+} \frac{\partial x'^i}{\partial x^j} \mid_{(x_0,y_0\pm \epsilon)} = \mp \delta^i_j.
\end{equation}
Applying (\ref{coordinateChange}), we can immediately see that 
the boundary conditions for $A_\mu'$ acquire an additional minus sign:
\begin{equation}
\label{newBC}
A_\mu'(\pm x',y_0') = -\sigma_1 A_\mu'(\mp x',y_0') \sigma_1.
\end{equation}

Let us now study the cases $n = 1$ and $n = 2$ for $M_0 = S^2$ before attempting to generalize
to arbitrary $n$.

\begin{figure}[tb]
\centerline{
\includegraphics[scale=0.5]{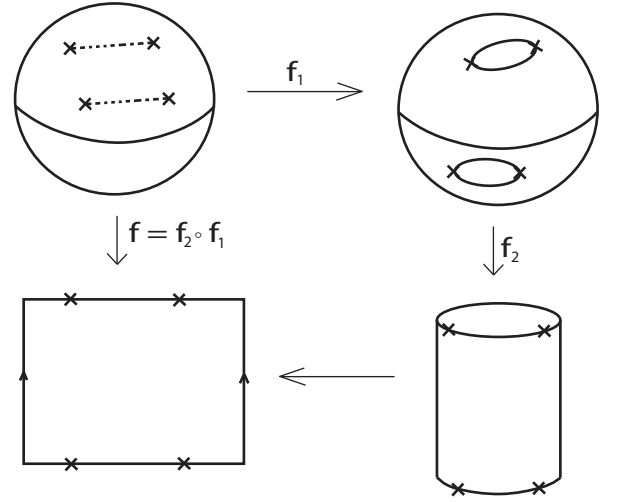}
}
\caption{
\label{diffeoFigure2}
Two pairs of $Z_2$ vortices on a sphere. This sequence of diffemorphisms illustrates that this situation
is equivalent to $M$ being a cylinder.}
\end{figure}

We begin by considering the case $n = 2$, the case of two pairs 
of $Z_2$ vortices on a sphere. Consider also the diffeomorphism $f$
shown in Figure \ref{diffeoFigure2}. Clearly, the situation with
two pairs of $Z_2$ vortices on a sphere is equivalent to having the gauge field $A_\mu$ 
defined on the space 
\begin{equation}
M = \{(x,y) \: | \: 0 \leq y \leq L, x \in \mathbb{R} \: \% \: L\},
\end{equation}
for any $L$, with the following periodicity/boundary conditions:
\begin{align}
\label{bcn2}
A_\mu(x+L,y) = A_\mu(x,y),
\nonumber \\
A_\mu(x,L) = - \sigma_1 A_\mu (-x,L) \sigma_1,
\nonumber \\
A_\mu(x,0) = - \sigma_1 A_\mu(-x,0) \sigma_1,
\end{align}
and with the action of (\ref{lagrangian}).
We can now define a new, continuous field $c_{\mu}$ defined on
\begin{equation}
\tilde{M} = \{ (x,y) | x \in \mathbb{R} \: \% \: L, y \in \mathbb{R} \: \% \: 2L \}
\end{equation}
as follows:
\begin{equation}
c_{\mu}(x,y) = \left\{ 
  \begin{array}{ll}
    a_{\mu}(x,y) &  0 \leq y \leq L \\
    -\tilde{a}_{\mu}(-x,2L - y) &  L \leq y \leq 2L
  \end{array} \right.
\end{equation}
where now $c_\mu$ is doubly periodic:
\begin{equation}
c_{\mu}(x,y) = c_{\mu}(x+L,y) = c_{\mu}(x,y+2L).
\end{equation}

Recall that the $U(1) \times U(1)$ gauge transformations on $A_\mu$ are of the form 
\begin{align}
U & = \left(\begin {array}{cc}
    e^{if}  & 0 \\
    0  & e^{ig}\\
\end{array}
\right)
\nonumber \\
A_\mu & \rightarrow A_\mu + i U \partial_\mu U^{-1}.
\end{align}
These gauge transformations must preserve the boundary conditions (\ref{bcn2}) on $A_\mu$. This implies
that $U$ obeys the following boundary conditions:
\begin{align}
U(x+L,y) = U(x,y),
\nonumber \\
U(x,L) = \sigma_1 U^{-1} (-x,L) \sigma_1,
\nonumber \\
U(x,0) = \sigma_1 U^{-1}(-x,0) \sigma_1.
\end{align}
Just as we defined $c_\mu$ from $A_\mu$, we can define the gauge transformation 
that acts on $c_\mu$ in the following way:
\begin{equation}
h(x,y) = \left\{ 
  \begin{array}{ll}
    f(x,y) &  0 \leq y \leq L \\
    -g(-x,2L - y) &  L \leq y \leq 2L
  \end{array} \right.
\end{equation} 
so that the gauge transformation $U$ acts on $c_\mu$ as:
\begin{equation}
c_\mu \rightarrow c_\mu - \partial_\mu h
\end{equation}
So we see that $c_\mu$ behaves like a typical $U(1)$ gauge 
field defined on a torus. In particular, the only condition
on $h(x,y)$ is that $e^{ih(x,y)}$ be doubly periodic, which 
allows for the possibility of large gauge transformations along
the two non-contractible loops of the torus. 

In the $A_0 = 0$ gauge, the Lagrangian can be written as:
\begin{align}
L = \epsilon^{ji} \int d^2x \: [\frac{k}{4 \pi} (a_i \dot{a}_j + \tilde{a}_i \dot{\tilde{a_j}}) 
+ \frac{k-l}{4\pi} (a_i \dot{\tilde{a}}_j + \tilde{a}_i\dot{a_j})],
\end{align}
where the integration is over the region $0 \leq x,y \leq L$. 
In terms of $c_\mu$:
\begin{align}
\int_0^L dx \int_0^L dy \: (a_i \dot{a}_j + \tilde{a}_i \dot{\tilde{a_j}}) = \int_0^L dx \int_0^{2L} dy \: c_i \dot{c}_j.
\end{align}
Using $\tilde{a}_j(x,y) = - c_j(-x,2L-y)$, we see:
\begin{align}
&\int_0^L \int_0^L d^2x  \: a_i \dot{\tilde{a}}_j = - \int_0^L \int_0^L d^2x \: c_i(x,y) \dot{c}_j(-x,2L-y),
\nonumber \\
&\int_0^L \int_0^L d^2x \: \tilde{a}_i \dot{a}_j = -\int_0^L dx \int_L^{2L} dy \: c_i(x,y) \dot{c}_j(-x,2L-y).
\end{align}
Therefore we can write the action in terms of $c_\mu$ as:
\begin{align}
\label{cAction}
L = \epsilon^{ji} \int_0^L dx \int_0^{2L} dy \:  [\frac{k}{4\pi} c_i \dot{c}_j - \frac{k-l}{4\pi} c_i\dot{c}_j(-x,2L-y)].
\end{align}
The equation of motion for $c_0$ serves as a constraint for zero field strength, which implies that
we can parameterize $c_i$ as
\begin{align}
\label{decomp}
c_i(x,y,t) = \frac{2\pi}{L_i} X_i(t) + \tilde{c}_i(x,y,t).
\end{align}
The large gauge transformations take $X_i \rightarrow X_i + integer$. The topological degeneracy is 
given by the degeneracy of this zero-mode sector. The action of the zero-mode sector is found upon substituting 
(\ref{decomp}) into the action (\ref{cAction}):
\begin{equation}
L = 2 \pi l X_2 \dot{X}_1.
\end{equation}
Now we must make sure that we project onto the $Z_2$ invariant sector. 
The $Z_2$ exchanges $a$ and $\tilde{a}$, so it takes $c(x,y) \rightarrow -c(x,y+L)$ if $y \leq L$
and $c(x,y) \rightarrow -c(x,y-L)$ if $y \geq L$. Thus, the action of the $Z_2$ is to take
the zero-modes to minus themselves: $X_i \rightarrow -X_i$. The states can be labelled by 
$|n \rangle$, where $n$ is an integer and with the identifications $|n \rangle = |n + l\rangle$.
Thus, before the projection, there are $l$ states. If $l$ is even, then there are two fixed points of the
of the $Z_2$ action, so in all there are $l/2 + 1$ $Z_2$ invariant states. If $l$ is odd, there are
only $(l+1)/2$ $Z_2$ invariant states. 

\begin{figure}[tb]
\centerline{
\includegraphics[scale=0.5]{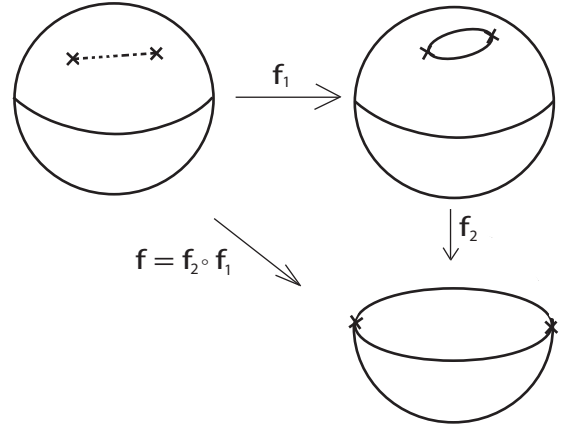}
}
\caption{\label{diffeo1}
A single pair of $Z_2$ vortices on a sphere. This sequence of diffemorphisms illustrates that this situation
is equivalent to $M$ being a hemisphere, but with a different set of boundary conditions on $A_\mu$.}
\end{figure}

Consider now the case of a single pair of $Z_2$ vortices on a sphere and 
the diffeomorphism $f$ shown in Fig. \ref{diffeo1}. Clearly, the situation
with a single pair of $Z_2$ vortices is equivalent to having the gauge field
$A_\mu$ defined on a hemisphere, but with modified boundary conditions on the $A_\mu$.
Let the angular coordinates $(\theta, \varphi)$ be defined so that the locations
of the two $Z_2$ vortices are $(\pi/2, 0)$ and $(\pi/2, \pi)$ for the left
and right vortices, respectively. The south pole is at $\theta = \pi$. 
As in the previous case with two $Z_2$ vortices, the boundary conditions
on $A_\mu$ at $\theta = \pi/2$ are as follows:
\begin{equation}
A_\mu(\pi/2,\varphi) = -\sigma_1 A_\mu(\pi/2, -\varphi) \sigma_1
\end{equation}
As a result, we can define a new, continuous gauge field $c_\mu$ on a sphere as follows:
\begin{equation}
c_\mu(\theta,\varphi) =  \left\{ 
  \begin{array}{ll}
    a_{\mu}(\theta,\varphi) &  \pi/2 \leq \theta \leq \pi \\
    -\tilde{a}_{\mu}(\pi-\theta,-\varphi) &  0 \leq \theta \leq \pi/2
  \end{array} \right.
\end{equation}
It is easy to see that in this case, there is no possibility for large gauge transformations
or holonomies around non-contractible loops. The Lagrangian will be given 
by an expression similar to (\ref{cAction}), but this time the degeneracy will be 1. 

\begin{figure}[thp]

\centerline{
\includegraphics[scale=0.45]{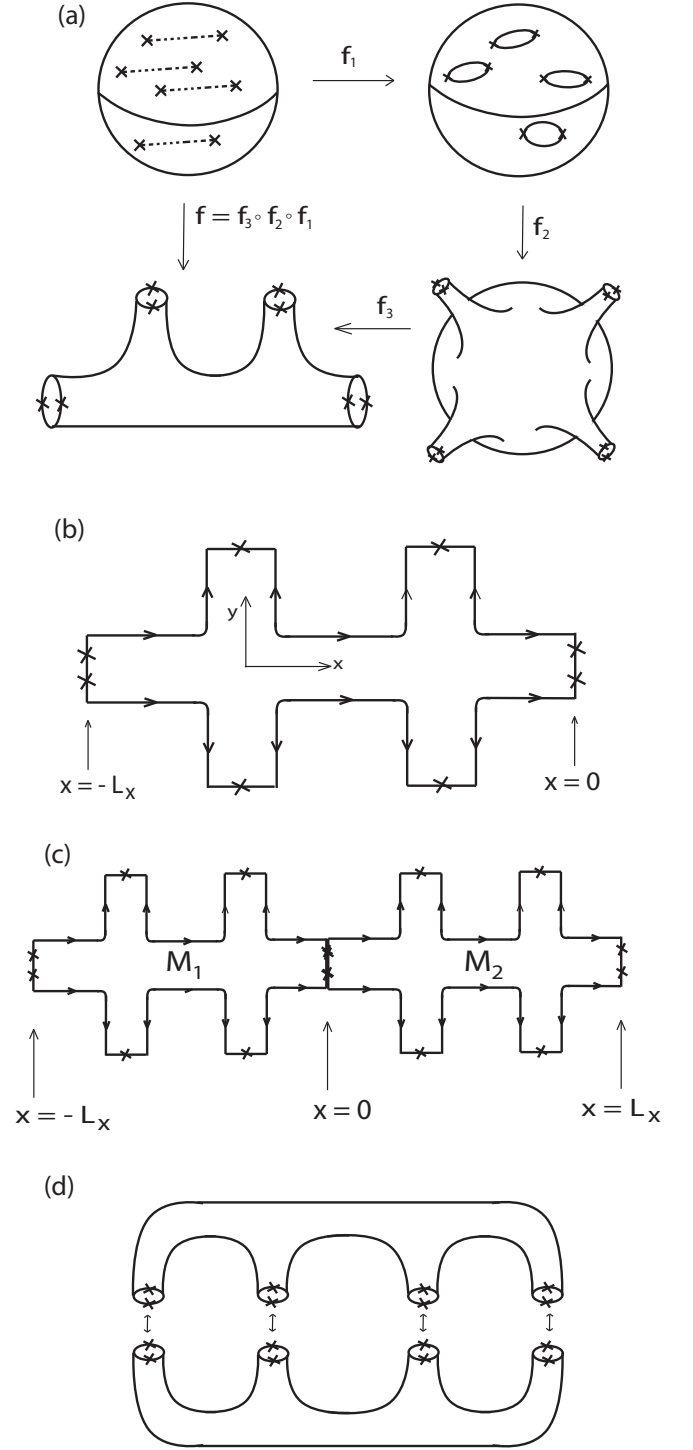}
}
\caption{
\label{diffeoN}
Four pairs of $Z_2$ vortices on a sphere. (a) This sequence of diffeomorphisms shows that we can
think of the situation with this many vortices as a gauge field defined on the surface shown in
the lower left figure, which looks like half of a genus $g = 3$ surface. (b) The figure in the lower
left of (a) can be cut open as shown here. The arrows on the figure indicate how the points on the boundaries
should be identified. (c) Two copies of the figure in (b). (d) Gluing together two copies along their
boundaries gives a genus $g=3$ surface. For general $n$, this procedure gives a surface of genus $g = n-1$.  }
\end{figure}

We now tackle the case for general $n$. Suppose that there
are $n$ pairs of $Z_2$ vortices on a sphere. We will define
the new gauge field $c_\mu$ on a genus $g = (n-1)$ surface 
in the following way. From Figure \ref{diffeoN}, we can 
clearly see that the situation with four pairs of vortices
is equivalent to having a gauge field $A_\mu$ defined on the 
surface shown in the lower left of Fig \ref{diffeoN}a and 
Fig. \ref{diffeoN}b, with modified boundary conditions. 
The generalization from four to $n$ is obvious. Consider the 
space shown in Figure \ref{diffeoN}c, which contains two copies 
of the original space. Parametrize this doubled space with the coordinates 
$\vec r = (x,y)$. We will refer to the copy on the left side, which 
has $x \leq 0$, as $M_1$; the copy on the right side, which has $x \geq 0$, 
will be referred to as $M_2$. Suppose that the length in the $x$ direction
of each copy is $L_x$, so that the total horizontal length of the doubled space is 
$2 L_x$. 

Consider a map $R$ defined on this doubled space with the following 
properties: $R$ takes $M_1$ to $M_2$ and $M_2$ into $M_1$
in such a way that $R \circ R = 1$, it has unit Jacobian, and it maps the
boundaries of $M_1$ and $M_2$ into each other. The way it maps $\partial M_1$
and $\partial M_2$ into each other is illustrated in Figure \ref{diffeoN}d;
if we identify $\partial M_1$ and $\partial M_2$ using the map $R$,
then we obtain a surface of genus $g = n - 1$, which we call $M$. In the coordinates
illustrated in Figure \ref{diffeoN}c, this way of mapping $\partial M_1$ and
$\partial M_2$ results in the following boundary conditions on $A_\mu$: 
\begin{equation}
\label{newBCn}
A_\mu(x,y) = - \sigma_1 A_\mu(R^1(x,y) - L_x, R^2(x,y)) \sigma_1,
\end{equation}
for $(x,y) \in \partial M_1$ and where $R^i(x,y)$ is the $i$th
coordinate of $R$ (note that $A_\mu(x,y)$ is only defined for $-L_x \leq x \leq 0$).
This allows us to define a continuous
gauge field $c_\mu$, defined on the doubled space $M$, in the following way:
\begin{equation}
\label{cDefn}
c_\mu(x,y) =  \left\{ 
  \begin{array}{ll}
    a_{\mu}(x,y) &  x \leq 0 \\
    -\tilde{a}_{\mu}(R(x,y)) &  x \geq 0 
  \end{array} \right.
\end{equation}

We now rewrite the various terms in the action in terms of $c_\mu$.
\begin{align}
\int_{M_1} d^2x \: \tilde{a}_i \dot{\tilde{a}}_j &= \int_{M_2} d^2x \: \tilde{a}_i(R(x,y)) \dot{\tilde{a}}_j(R(x,y)) 
\nonumber \\
 &= \int_{M_2} d^2x \: c_i \dot{c}_j 
\end{align}
The cross terms give a nonlocal term in the action:
\begin{align}
\int_{M_1} d^2x \: a_i \dot{\tilde{a}}_j = -\int_{M_1} d^2x \: c_i (x,y) \dot{c}_j(R(x,y))
\nonumber \\
\int_{M_1} d^2x \: \tilde{a}_i \dot{a}_j = -\int_{M_2} d^2x \: c_i (x,y) \dot{c}_j(R(x,y))
\end{align}
Thus the Lagrangian is:
\begin{equation}
L = \epsilon^{ji} \int_{M} d^2x \: [ \frac{k}{4 \pi} c_i \dot{c}_j - \frac{k-l}{4\pi} c_i(x,y) \dot{c}_j(R(x,y)) ] .
\end{equation}
As usual in pure Chern-Simons theory, the equation of motion for $c_0$ implies that
the gauge field must be flat. It is therefore characterized by the value of 
$\oint_C c \cdot dl$ along its non-contractible loops. To parametrize the gauge
field, as is typical we introduce a canonical homology basis $\alpha_i$ and $\beta_i$
such that the $\alpha_i$ $(\beta_i)$ do not intersect while $\alpha_i$ and $\beta_j$
intersect if $i = j$. Then we introduce the dual basis $\omega_i$ and $\eta_j$,
which satisfy:
\begin{align}
\int_{\alpha_i} \omega_j = \delta_{ij} \;\;\;\; \int_{\beta_i} \omega_j = 0,
\nonumber \\
\int_{\alpha_i} \eta_j = 0 \;\;\;\; \int_{\beta_i} \eta_j = \delta_{ij}.
\end{align}
Since $c$ must be a flat connection, we can parametrize it as
\begin{equation}
c = c_1 dx^1 + c_2 dx^2 = 2\pi (x^i \omega_i + y^i \eta_i).
\end{equation}
Two connections $c$ and $c'$ are gauge-equivalent if 
\begin{equation}
x'^i - x^i = \text{ integer }, \;\;\;\; y'^i - y^i = \text{ integer }.
\end{equation}
Notice that here, the action of $R$ is trivial on the canonical homology cycles.
This is because of the way the genus $n-1$ surface was glued together
from its pieces (see Figure \ref{diffeoN}d).
This is in contrast to eqn. (\ref{sheetExchangeAction}), which we obtained 
when we were analyzing the ground state degeneracy on higher genus surfaces.
Therefore, the action in terms of the $x^i$ and $y^i$ becomes simply
\begin{equation}
L = 2 \pi l x^i \dot{y}^i,
\end{equation} 
for $i = 1, \cdots, n-1$.
However, the $Z_2$ action here is not exactly the same as the action of the
sheet exchange map $R$. This is because the $Z_2$ exchanges $a$ and $\tilde{a}$,
so it takes $c(x,y) \rightarrow -c(R(x,y))$. Thus, the action of the $Z_2$
is to change the sign of the $x^i$ and $y^i$: $x^i \rightarrow -x^i$
and $y^i \rightarrow -y^i$ for every $i$, under the action of the $Z_2$.
Before projection, it is clear that we have $l^{n-1}$ states. These
can be labelled in the following way
\begin{equation}
\label{Z2vortexStates}
\otimes_{i=1}^{n-1} |m_i \rangle,
\end{equation}
where $m_i$ is an integer and $m_i \sim m_i + l$. The $Z_2$ action
takes $m_i \rightarrow -m_i$. So if $l$ is odd, there is one state
that is already $Z_2$ invariant: the state with $m_i = 0$ for all $i$.
There are $l^{n-1} -1$ remaining states, and exactly half of them are
$Z_2$ invariant. Thus if $l$ is odd, the degeneracy of ($Z_2$-invariant) 
states in the presence of $n$ pairs of $Z_2$ vortices on a sphere is
$(l^{n-1} + 1)/2$. For $l = 2$, $|m_i \rangle = |0 \rangle$ or $|1 \rangle$,
which are both $Z_2$ invariant, so for $l=2$ the degeneracy in the presence of $n$
pairs of $Z_2$ vortices on a sphere is $2^{n-1}$. One may ask also
about the number of states that are not $Z_2$ invariant. These
may correspond to a different set of quasiparticle states that carry $Z_2$ charge.
We see that there are $(3^{n-1} - 1)/2$ $Z_2$ non-invariant states for $l = 3$ if 
there are $n$ pairs of $Z_2$ vortices on a sphere. 

\begin{table}
\begin{tabular}{|c|c|c|}
\hline
No. $Z_2$ vortex pairs & No. $Z_2$ inv. states & No. $Z_2$ non-inv. states \\
\hline
$n$ & $(3^{n-1} + 1)/2$ &  $(3^{n-1} - 1)/2$ \\
\hline
\hline
$1$ & $1$ & $0$ \\
\hline
$2$ & $2$ & $1$ \\
\hline
$3$ & $5$ & $4$ \\
\hline
$4$ & $14$ & $13$ \\
\hline
$5$ & $41$ & $40$\\
\hline
$6$ & $122$ & $121$\\
\hline
\end{tabular}
\caption{
\label{Z2vortexDegTable}
Some values of the $Z_2$ vortex degeneracy for $l = 3$ for the
$Z_2$ invariant states, given by $(3^{n-1} + 1)/2$,
and for the $Z_2$ non-invariant states $(3^{n-1} - 1)/2$.}
\end{table}

\subsection{Comparison to Quasiparticles in $Z_4$ Parafermion and Pfaffian FQH states }

Let us now compare the results from the previous section to the quasiparticles
in the Pfaffian and $Z_4$ parafermion FQH states.

The topological properties of the quasiparticles in FQH states can be computed
through the pattern of zeros approach\cite{BW0932, WW0808, WW0809}  
or through their connection to conformal field theory.\cite{MR9162,RR9984}
In the Pfaffian quantum Hall state, there are two main types of 
quasiparticles, corresponding to two different representations of a magnetic
translation algebra.\cite{BW0932} These two classes of quasiparticles are commonly labelled 
in the following way:
\begin{equation}
\psi e^{i Q\frac{1}{\nu} \varphi},  \;\;\; \sigma e^{i Q \frac{1}{\sqrt{\nu}}\varphi} ,
\end{equation}
where $\psi$ is the Majorana fermion and $\sigma$ is the spin field
of the Ising CFT. $Q$ is the charge of the quasiparticle and $\nu$ is the filling fraction of
the quantum Hall state. The ones made of $\psi$ are Abelian; there are $2q$ of them
when the filling fraction is $\nu = 1/q$. The ones made of $\sigma$ are non-Abelian;
there are $q$ of them and their quantum dimension is $\sqrt{2}$.
In the presence of $n$ pairs of the $\sigma$ quasiparticles, the
Pfaffian state has a degeneracy of $2^{n-1}$ on a sphere. This follows from the fusion rules
of the conformal primary fields in the Ising CFT:
\begin{align}
\psi \psi &= 1
\nonumber \\
\sigma \sigma &= 1 + \psi
\nonumber \\
\psi \sigma &= \sigma.
\end{align}

Similarly, the quasiparticles of the $Z_4$ parafermion state compose three different representations
of a magnetic translation algebra, and these three classes are commonly labelled  as
\begin{equation}
e^{i Q \sqrt{1/\nu} \varphi}, \;\;\; \Phi^1_1 e^{i Q \sqrt{1/\nu}\varphi}, \;\;\; \Phi^2_2 e^{i Q \sqrt{1/\nu}\varphi}.
\end{equation}
When the filling fraction is $\nu = \frac{2}{2k-3}$, there are $2(2k-3)$ Abelian quasiparticles,
$2(2k-3)$ of the $\Phi^1_1$ quasiparticles and $2k-3$ of the $\Phi^2_2$ quasiparticles.
The $\Phi^1_1$ quasiparticles have quantum dimension $\sqrt{3}$ and the $\Phi^2_2$ quasiparticles
have quantum dimension $2$. 

The parafermionic primary fields in the $Z_4$ parafermion CFT have the fusion rules:
\begin{align}
\Phi^0_m \times \Phi^l_{m'} &= \Phi^l_{m+m'}
\nonumber \\
\Phi^1_1 \times \Phi^1_1 &= \Phi^2_2 + \Phi^0_2,
\nonumber \\
\Phi^2_2 \times \Phi^2_2 &= \Phi^0_0 + \Phi^0_4 + \Phi^2_0,
\nonumber \\ 
\Phi^1_1 \times \Phi^2_2 &= \Phi^1_{-1} + \Phi^1_3.
\end{align}
The fusion rules imply:
\begin{align}
\label{phi11fusion}
\Phi^1_1 \Phi^1_1 &= \Phi^2_2 + \Phi^0_2
\nonumber \\
(\Phi^1_1 \Phi^1_1 )^2 &= \Phi^0_0 + 2\Phi^0_4 + 3\Phi^2_0 
\nonumber \\
(\Phi^1_1 \Phi^1_1 )^3  &= 9\Phi^2_2  + 4\Phi^0_2 + 5 \Phi^0_6
\nonumber \\
(\Phi^1_1 \Phi^1_1)^4  &= 27 \Phi^2_0 + 13 \Phi^0_4 + 14 \Phi^0_0 
\nonumber \\
(\Phi^1_1 \Phi^1_1)^5  &= 81 \Phi^2_2 + 40 \Phi^0_6 + 41 \Phi^0_2
\nonumber \\
(\Phi^1_1 \Phi^1_1)^6  &= 243 \Phi^2_0 + 122 \Phi^0_4 + 121 \Phi^0_0
\end{align}
There appears to be a connection between the $\Phi^1_1$ quasiparticles and the $Z_2$ vortices.
First, notice that one member of a pair of $Z_2$ vortices should be conjugate
to the other member. This is because a pair of $Z_2$ vortices can be created out of the
vacuum on a sphere. Suppose that we identify one member of a pair with the operator 
$V_\sigma = \Phi^1_1 e^{i Q \frac{1}{\sqrt{\nu}}\varphi}$ and the other member with
its conjugate $V_{\bar \sigma} = \Phi^1_{-1} e^{-i Q \frac{1}{\sqrt{\nu}}\varphi}$.
From eqn. \ref{phi11fusion}, we see that the number of ways to fuse 
to the identity for $(V_\sigma V_{\bar\sigma})^n = \Phi^0_{-2n} (\Phi^1_1 \Phi^1_1)^n$
is as displayed in Table \ref{Phi11qpFusion}. Notice that this agrees exactly
with the number of $Z_2$ invariant states for $n$ $Z_2$ vortices on a sphere
(see Table \ref{Z2vortexDegTable})!
\begin{table}
\begin{tabular}{|c|c|c|}
\hline
$n$ & No. of ways to fuse to $\Phi^0_0 = 1$ & No. of ways to fuse to $\Phi^0_4$\\
\hline
$1$ & $1$ & $0$ \\
\hline
$2$ & $2$ & $1$ \\
\hline
$3$ & $5$ & $4$\\
\hline
$4$ & $14$ & $13$ \\
\hline
$5$ & $41$ & $40$\\
\hline
$6$ & $122$ & $121$ \\
\hline
\end{tabular}
\caption{
\label{Phi11qpFusion}
Number of ways of fusing to the identity or to $\Phi^0_4$ for the fusion of
$(V_\sigma V_{\bar\sigma})^n$, where $V_\sigma = \Phi^1_1 e^{i Q \frac{1}{\sqrt{\nu} \varphi}}$
and $V_{\bar \sigma} = \Phi^1_{-1} e^{-i Q \frac{1}{\sqrt{\nu} \varphi}}$.
}
\end{table}

Notice that the number of ways for $(V_\sigma V_{\bar \sigma})^n$ to fuse
to the quasiparticle $\Phi^0_4$ is exactly equal to the number of 
$Z_2$ non-invariant states that we obtain from $n$ pairs of $Z_2$ vortices
(see Table  \ref{Phi11qpFusion} and \ref{Z2vortexDegTable})!
This shows that the $Z_2$ non-invariant states have a meaning as well. 
These states carry non-trivial $Z_2$ charge, so we interpret this as a situation in which 
there are $n$ pairs of $Z_2$ vortices and an extra $Z_2$ charged quasiparticle. 
The above fusion indicates that we should associate this $Z_2$ charged quasiparticle
to the operator $\Phi^0_4$. 

Based on this quantitative agreement between the properties of the $Z_2$
vortices and results from the $Z_4$ parafermion FQH state, we conclude 
that for a pair of $Z_2$ vortices, one of them should be associated 
with an operator of the form $\Phi^1_1 e^{i Q \frac{1}{\sqrt{\nu}} \varphi}$ and the one
to which it is connected by a branch cut should be associated with
$\Phi^1_{-1} e^{-i Q \frac{1}{\sqrt{\nu}} \varphi}$. Furthermore, the
possibility of $Z_2$ non-invariant states should be interpreted
as the possibility for the $Z_2$ vortices to fuse to an electromagnetically neutral 
$Z_2$ charged quasiparticle, which we associate with the operator $\Phi^0_4$.

We have not seen how to understand the quantization of electromagnetic charge, $Q$,
for the $Z_2$ vortices. The external electromagnetic field couples to the field
$a^+ = a + \tilde{a}$, so we expect electromagnetically charged quasiparticles
to carry flux of the $a^+$ field. The quantization of charge for the quasiparticles 
generally arises from the constraint that quasiparticles are mutually local with respect
to electrons. We should be able to see how the $Z_2$ vortices
must carry certain quantized units of $a^+$ flux, but we have not performed this analysis.  

\section{Conclusion}

In this paper, we have computed several topological properties of $U(1)
\times U(1) \rtimes Z_2$ Chern-Simons theory and discussed its
relation to the Pfaffian and $Z_4$ parafermion FQH states. For the $l
= 3$ $U(1) \times U(1) \rtimes Z_2$ Chern-Simons theory, many
topological properties agree with those of the $Z_4$ parafermion state, which
strongly suggests that the Chern-Simons theory correctly describes all
of the topological properties of this state. This identification also
suggests that the phase transition between the $(k, k, k-3)$ bilayer
state and the $Z_4$ parafermion FQH state can be continuous and may,
for instance, be described by a $Z_2$ transition in 2+1 dimensions.
In the simplest case, for $k = 3$, this would be a continuous $Z_2$
transition at $\nu = 2/3$ between the $(3,3,0)$ state and the 
non-Abelian $Z_4$ parafermion state. We leave a study of the phase 
transition itself for future work. 

More generally, the methods in this paper may be extended to compute
topological properties of Chern-Simons theories with disconnected
gauge groups of the form $G \rtimes H$, where $G$ is a connected Lie
group and $H$ is a discrete automorphism group of $G$. There may be
other situations also in which an $n$-layer FQH state passes through a
phase transition to an $m$-layer FQH state, where the Chern-Simons
gauge theories for each of the phases will be $G \rtimes H_n$ and $G
\rtimes H_m$, respectively, and the phase transition will be described
by a discrete gauge symmetry-breaking of $H_n$ to $H_m$. We expect
that such a scenario may be possible if the central charges of the
corresponding edge theories are the same for the two phases. In this
paper, for example, we found that even though there is a phase transition
between the $(k,k,k-2)$ bilayer states and the Pfaffian states as the
interlayer tunneling is increased, the $l=2$ $U(1) \times U(1) \rtimes Z_2$
theory does not describe the Pfaffian state. In contrast, there is a possible
phase transition between the $(k,k,k-3)$ bilayer states and the $Z_4$ parafermion
states, and in this case the $l=3$ $U(1) \times U(1) \rtimes Z_2$ theory 
does correctly describe the $Z_4$ parafermion state. One
way to understand why simply gauging a $Z_2$ symmetry does 
not describe the Pfaffian state is that the central
charges of the edge theory changes as the interlayer tunnelling is
tuned through a phase transition from the bilayer $(k, k, k-2)$ phase
to the Pfaffian state, which indicates there is additional physics
taking place that this approach does not capture here. The parent bilayer
Abelian phase has $c = 2$, as does the edge theory of the 
$Z_4$ parafermion state, while the edge theory of the Pfaffian 
state has $c = 3/2$. 

We would like to thank Brian Swingle for helpful discussions. 
This research is supported by NSF Grant No. DMR-0706078.

\appendix

\section{ $Z_4$ Parafermion FQH states and Projective Construction }
\label{projConstruction}

Here we explain, from the point of view of a procedure called 
projective construction,\cite{W9927} how to understand that 
the $(k, k, k-3)$ bilayer wave function, upon symmetrization, yields 
the $Z_4$ parafermion wave function at $\nu = \frac{2}{2k - 3}$, 
and a different explanation for why we expect that the corresponding Chern-Simons theory 
should have the gauge group $U(1) \times U(1) \rtimes Z_2$. 

In the projective construction approach, one writes the electron operator (which is either 
bosonic or fermionic, depending on whether we are interested in FQH states of bosons or
fermions) in terms of several other fermionic fields, $\psi_1, \cdots \psi_n$, referred to as ``partons:''  
\begin{equation}
\Psi_e = \psi_{\alpha_1} \cdots \psi_{\alpha_n} C_{\alpha_1 \cdots \alpha_n},
\end{equation}
where $C_{\alpha_1 \cdots \alpha_n}$ are constant coefficients. 
The continuum field theory that describes interacting electrons in an external 
magnetic field can then be rewritten in terms of the partons and a gauge field. 
The introduction of the partons expands the Hilbert space, so the gauge field 
is included in order to project the states onto the physical Hilbert space, which
is generated by the electron operator.  If the partons form a state
$|\Phi_{parton} \rangle$, the electron wave function is the projection onto the
physical electronic Hilbert space: 
\begin{equation}
\label{wfn}
\Phi_e(z_1, \cdots, z_N) = \langle 0 | \prod_i \Psi_{e}(z_i) | \Phi_{parton} \rangle.
\end{equation}


If $G$ is the group of transformations on the partons that keeps the electron operator
invariant, then the continuum field theory description will be partons interacting
with a gauge field with gauge group $G$, which ensures that physical excitations,
which are created by electron operators, will be singlets of the group $G$. Since
the partons are assumed to form a gapped integer quantum Hall state, they can be integrated out
to obtain a Chern-Simons theory with gauge group $G$. 

For example, if we choose the electron operator to be
\begin{align}
\Psi_{e;1/3} = \psi_1 \psi_2 \psi_3, 
\end{align}
then $\Psi_{e;1/3}$ is an $SU(3)$ singlet.  If we assume that the
partons each form a $\nu = 1$ integer quantum Hall state, then the
electron wave function is
\begin{equation}
\Phi_{e;1/3}(z_1, \cdots, z_N) = \prod_{i<j} (z_i - z_j)^3,
\end{equation}
which is the Laughlin $\nu = 1/3$ wave function. The continuum field
theory is a theory of three fermions coupled to an $SU(3)$ gauge
field. Integrating out the partons will yield a $SU(3)_1$ Chern-Simons
theory. This theory is equivalent to the $U(1)_3$ Chern-Simons theory,
which is the topological field theory for the $\nu = 1/3$ Laughlin
state. 

If we choose the electron operator to be
\begin{align}
\Psi_{e;pf} = \psi_1 \psi_2 + \psi_3 \psi_4,
\end{align}
and assume the partons form a $\nu = 1$ IQH state, we can obtain the
wave function after projection by using the following observation. The
$\nu = 1$ wave functions are equal to free chiral fermion correlators
of a $1+1$-dimensional CFT:
\begin{align}
\Phi_{\nu = 1} &= \langle 0 | \prod_i \psi(z_i) | \nu = 1 \rangle = \prod_{i < j} (z_i - z_j)
\nonumber \\
&\sim \langle e^{-i N \phi(z_{\infty})} \prod_{i = 1}^N \psi(z_i) \rangle,
\end{align}
where in the first line, $\psi(z_i)$ is a free fermion operator that annihilates a 
fermion at position $z_i$ and $|\nu = 1 \rangle$ is the $\nu = 1$ integer quantum Hall state for
the fermion $\psi$; in the second line, $\psi(z_i)$ is interpreted as a 
free chiral fermion operator in a 1+1-d CFT and $\frac{1}{2\pi} \partial \phi = \psi^{\dagger} \psi$
is the density of the fermions. From this, it follows that the wave function (\ref{wfn}) with the electron operator
$\Psi_{e;pf}$ can be obtained by taking the correlator
\begin{equation}
\Phi_{e;pf} \sim \langle e^{-i N \phi(z_{\infty})} \prod_{i = 1}^N \Psi_{e;pf} (z_i) \rangle,
\end{equation}
where $\Psi_{e;pf} = \psi_1 \psi_2 + \psi_3 \psi_4$ and $\psi_i(z)$ is
now interpreted as a free complex chiral fermion in a 1+1d CFT. The operator
product algebra generated by the electron operator in this case can be
checked to be reproduced if we instead write the electron operator as 
\begin{equation}
\Psi_{e;pf} = \psi \eta, 
\end{equation}
where $\eta$ is a Majorana fermion and $\psi$ is a free chiral fermion. The correlation
function with $N$ insertions of this operator is known to yield the $\nu = 1$ Pfaffian wave function.
The gauge group that keeps $\Psi_{e;pf}$ invariant is $SO(5)$, and thus the Chern-Simons theory for the
$\nu = 1$ Pfaffian is a $SO(5)_1$ Chern-Simons theory.\cite{W9927}

Now consider a bilayer wave function, where we have two electron operators, one for each
layer, and the wave function is given by:
\begin{equation}
\Phi(\{z_i\}, \{w_i \}) \sim \langle e^{-i N \phi(z_{\infty})} \prod_{i = 1}^N \Psi_{e1} (z_i) \Psi_{e2} (w_i) \rangle.
\end{equation}
The single-layer wave function that can be obtained by symmetrizing or anti-symmetrizing over the
electron coordinates in the two layers can be obtained by choosing the single-layer electron operator
to be $\Psi_e = \Psi_{e1} + \Psi_{e2}$:
\begin{align}
\Phi( \{z_i\}) &= S\{ \Phi( \{z_i\}, \{w_i\}) \} 
\nonumber \\
&\sim \langle e^{-i N \phi(z_{\infty})} \prod_{i = 1}^{2N} (\Psi_{e1} (z_i) + \Psi_{e2} (z_i)) \rangle,
\end{align}
where we have set $z_{N+i} = w_i$. 

In the case of the Pfaffian, this shows us that the $(2,2,0)$ state, when symmetrized, yields
the Pfaffian wave function. If we instead consider $\Psi_{e1} = \psi_1 \psi_2 \psi_3$ and
$\Psi_{e2} = \psi_4 \psi_5 \psi_6$, we obtain the $(3,3,0)$ state. The $(3,3,0)$ state, when symmetrized,
will therefore be given by 
\begin{equation}
\Phi( \{z_i\}) \sim \langle e^{-i N \phi(z_{\infty})} \prod_i \Psi_e(\{z_i\}) \rangle,
\end{equation}
with $\Psi_e = \psi_1 \psi_2 \psi_3 + \psi_4 \psi_5 \psi_6$.  It can be checked that
the operator product algebra generated by this electron operator is also generated by 
the operator $\Psi_e = \Phi_2^0 e^{i \sqrt{ 3/2} \phi}$, where $\Phi_2^0$ is a simple-current
operator in the $Z_4$ parafermion CFT and $\phi$ is a scalar boson. Thus, this wave function
is the wave function of the $Z_4$ parafermion FQH state at $\nu = 2/3$. Furthermore, 
the gauge group that keeps the electron operator invariant is $SU(3) \times SU(3) \rtimes Z_2$,
so we expect that the corresponding Chern-Simons theory for this phase should be
$SU(3)_1 \times SU(3)_1 \rtimes Z_2$ Chern-Simons theory, which we expect to be equivalent
to $U(1)_3 \times U(1)_3 \rtimes Z_2$ Chern-Simons theory. One would then guess that the 
generalization to the $(k, k, k-3)$ states and the $\nu = \frac{2}{2k-3}$ $Z_4$ parafermion
states is the $U(1) \times U(1) \rtimes Z_2$ Chern-Simons theory described in this paper. 

\section{More detailed discussion of the ground state degeneracy}
\label{torusDiscussion}

Here we like to discuss the the ground state degeneracy of the $U(1)
\times U(1) \rtimes Z_2$ Chern-Simons theory in more detail.  For
$l=2$, the filling fraction is $\nu = \frac{1}{k-1}$ and the formula
(\ref{torusDeg}) gives $9(k-1)$ states on a torus. Compare this to the
torus degeneracy of the $\nu = \frac{1}{k-1}$ Pfaffian state, which is
$3 (k-1)$. We see that the $U(1) \times U(1) \rtimes Z_2$ Chern-Simons
theory for $l=2$ has a torus ground state degeneracy that is three
times that of the Pfaffian state.  The origin of this factor of 3 can
be thought of in the following way. It is known that $O(2)_{2l}$
Chern-Simons theory has $l+7$ ground states\cite{MS8922} (see Appendix
\ref{Z2orbifold}). So, $U(1)_{k-1} \times O(2)_4$ has $9(k-1)$ ground
states on a torus. Furthermore, the gauge group $U(1) \times O(2)$ is
similar to $U(1) \times U(1) \rtimes Z_2$ if one considers the
positive and negative combinations of the two $U(1)$ gauge fields: if
one considers $a^+ = a + \tilde{a}$ and $a^- = a - \tilde{a}$, the
gauge group can be thought of as $U(1) \times O(2)$ because the action
of the $Z_2$ is to take $a^- \rightarrow - a^-$. Now, $O(2)$
Chern-Simons theory at level $2l$ is known to correspond to the $Z_2$
rational orbifold conformal field theory at level $2l$, which for
$l=2$, is known to be dual to two copies of the Ising
CFT.\cite{MS8922,DV8985} The Ising CFT has three primary fields, and
the CFT corresponding to the Pfaffian is one that contains an Ising
CFT and a $U(1)$ CFT. In this sense our theory has an extra copy of
the Ising model, which accounts for the extra factor of three in the
torus degeneracy. We can see this another way by noticing that the
central charge of the Ising CFT is 1/2 and the central charge of the
CFT that corresponds to the Pfaffian state is $c = 3/2$. Meanwhile,
the CFT corresponding to the $U(1) \times U(1) \rtimes Z_2$
Chern-Simons theory has $c = 2$, which corroborates the fact that it
has an extra copy of the Ising model.

For $l=3$, the filling fraction is $\nu = \frac{2}{2k-3}$ and
(\ref{torusDeg}) gives $5(2k-3)$ ground states on a torus. Compare
this to the $\nu = \frac{2}{2k-3}$ $Z_4$ parafermion state, which also
has a torus degeneracy of $5(2k-3)$.  This might be expected from the
fact that $O(2)_{2l}$ Chern-Simons theory corresponds to the $Z_4$
parafermion CFT when $l=3$. However, there is a crucial issue that needs to be addressed
here. In the case $l=2$, we could see that $U(1)_{k-1} \times O(2)_4$
Chern-Simons theory gives the same number of ground states on a torus
as the $U(1) \times U(1) \rtimes Z_2$ theory did, implying that we
could perhaps think of the $U(1)$ sector of the theory as separate
from the $O(2)$ sector. This fails in the $l = 3$ case. We
would be tempted to write $U(1)_{k-3/2} \times O(2)_6$, because $(k -
3/2) \times (3+7)$ gives the right ground state degeneracy. This fails
because the ground state degeneracy of $U(1)_{k-3/2}$ Chern-Simons
theory is not $(k-3/2)$. $U(1)_q$ Chern-Simons theory is typically
defined to have integer $q$, but the quantization procedure may also
be applied in cases where $q$ is not an integer. In these latter cases,
the quantum states do not transform as a one-dimensional representation
under large gauge transformations. One may wish to reject a theory
in which the quantum states are not gauge invariant, in which case 
$U(1)_q$ is not defined for non-integer $q$. On the other hand,
if these situations are allowed, then it can be shown that $U(1)_q$
Chern-Simons theory, for $q = p/p'$ (where $p$ and $p'$ are coprime),
has a torus degeneracy of $pp'$. \footnote{See \it e.g. \rm G. Dunne,
``Aspects of Chern-Simons Theory,'' from Les Houches Lectures 1998.}
Therefore, $U(1)_{k-3/2}$ Chern-Simons theory, to the extent that it is well-defined, 
has degeneracy $2(2k-3)$.  In either case, it is
clear that the $U(1)$ and $O(2)$ sectors cannot be disentangled and
that the correct definition of the theory is the $U(1) \times U(1)
\rtimes Z_2$ Chern-Simons theory presented here. 

To summarize, for $l=2$,  $U(1) \times U(1) \rtimes Z_2$ Chern-Simons
theory describes the Pfaffian state but with an extra copy of the
Ising model, while for $l=3$, the $U(1) \times U(1) \rtimes Z_2$ theory
gives the same ground state degeneracy as the $Z_4$ parafermion
quantum Hall state.

\section{$O(2)$ Chern-Simons Theory and $Z_2$ Rational Orbifold Conformal Field Theories}
\label{Z2orbifold}

Here we summarize previously known results from $O(2)$ Chern-Simons theory and
the $Z_2$ orbifold CFT and apply the $Z_2$ vortex analysis of this paper to the 
$O(2)$ Chern-Simons theory. 

Moore and Seiberg\cite{MS8922} first discussed Chern-Simons 
theories with disconnected gauge groups of the form $P \rtimes G$, where 
$G$ is a connected group with a discrete automorphism group $P$,
and the connection of these Chern-Simons theories to $G/P$ orbifold conformal field theories.
As a special example, they discussed the case where $G = U(1)$
and $P = Z_2$. In the 2d conformal field theory, this is known
as the $Z_2$ orbifold and it was explicitly analyzed in \Ref{DV8985}. 
It is the theory of a scalar boson $\varphi$ compactified
at a radius $R$, so that $\varphi \sim \varphi + 2\pi R$, and with an additional
$Z_2$ gauge symmetry: $\varphi \sim -\varphi$. 

\subsection{$Z_2$ Rational Orbifold CFT}

When $\frac{1}{2} R^2$ is rational, \it i.e. \rm $\frac{1}{2} R^2 = p/p'$, 
with $p$ and $p'$ coprime, then it is useful to
consider an algebra generated by the fields $j = i \partial \varphi$,
and $e^{\pm i \sqrt{2N} \varphi}$, for $N = p p'$. This algebra is referred to
as an extended chiral algebra. The infinite number of Virasoro primary fields
in the $U(1)$ CFT can now be organized into a finite number of representations 
of this extended algebra $\mathcal{A}$. There are $2N$ of these representations, and the
primary fields are written as $V_l = e^{il\varphi/\sqrt{2N}}$, with 
$l = 0, 1, \cdots, 2N-1$. 

In the $Z_2$ orbifold, one now considers representations of the smaller
algebra $\mathcal{A}/Z_2$. This includes the $Z_2$ invariant combinations of the
original primary fields, which are of the form $\cos(l \varphi/\sqrt{2N})$;
there are $N+1$ of these. In addition, there are 6 new operators. The gauging
of the $Z_2$ allows for twist operators that are not local with respect to the fields
in the algebra $\mathcal{A}/Z_2$, but rather local up to an element of $Z_2$. It turns
out that there are two of these twisted sectors, and each sector contains one field
that lies in the trivial representation  of the $Z_2$, and one field that lies
in the non-trivial representation of $Z_2$. These twist fields are labelled
$\sigma_1$, $\tau_1$, $\sigma_2$, and $\tau_2$. In addition to these, an in-depth
analysis\cite{DV8985} shows that the fixed points of the $Z_2$ action in the original $U(1)$ 
theory split into a $Z_2$ invariant and a non-invariant field. We have already 
counted the invariant ones in our $N+1$ invariant fields, which leaves 2 new
fields. In total, there are $N+7$ fields in the $Z_2$ rational
orbifold at ``level'' $2N$.

The dimension of the space of conformal blocks on a genus $g$ surface
is given by the following formula:\cite{V8860}
\begin{equation}
\text{dim }V_g = \text{Tr }\left( \sum_{i=0}^{N-1} N_i^2 \right)^{g-1} = \sum_{n=0}^{N-1} S_{n0}^{-2(g-1)}.
\end{equation}
The $S$ matrix was computed for the $Z_2$ orbifold in \Ref{DV8985}, so
we can immediately calculate the above quantity in this case. The result is:
\begin{align}
\label{dimVg}
\text{dim } V_g & = 2^{g-1} (2^{2g} + (2^{2g} - 1) N^{g-1} + N^g).
\end{align}

For $N = 2$, it was observed that the $Z_2$ orbifold is equivalent 
to two copies of the Ising CFT. For $N = 3$, it was observed that the
$Z_2$ orbifold is equivalent to the $Z_4$ parafermion CFT of Zamolodchikov
and Fateev.\cite{ZF8515}

In Tables \ref{Z2fieldsN2} and \ref{Z2fieldsN3} we list the
fields from the $Z_2$ orbifold for $N=2$ and $N=3$, their
scaling dimensions, and the fields in the $Ising^2$ or $Z_4$ 
parafermion CFTs that they correspond to.

\begin{table}
\begin{tabular}{|c|c|c|}
\hline
$Z_2$ Orb. field & Scaling Dimension, $h$ & $Ising^2$ fields \\
\hline
\hline
$1$ & $0$ & $\mathbb{I} \otimes \mathbb{I}$ \\
\hline
$j$ & $1$ & $\psi \otimes \psi$ \\
\hline
$\phi^1_N$ & $1/2$ & $\mathbb{I} \otimes \psi$ \\
\hline
$\phi^2_N$ & $1/2$ & $\psi \otimes \mathbb{I}$ \\
\hline
$\phi_1$ & $1/8$ & $\sigma \otimes \sigma$ \\
\hline
$\sigma_1$ & $1/16$ & $\sigma \times \mathbb{I}$ \\
\hline
$\sigma_2$ & $1/16$ & $\mathbb{I} \otimes \sigma$ \\
\hline
$\tau_1$ & $9/16$ & $\sigma \otimes \psi$ \\
\hline
$\tau_2$ & $9/16$ & $\psi \otimes \sigma$ \\
\hline
\end{tabular}
\caption{
\label{Z2fieldsN2}
Primary fields in the $Z_2$ orbifold for $N=2$, their 
scaling dimensions, and the fields from Ising$^2$ to which they
correspond.}
\end{table}

\begin{table}
\begin{tabular}{|c|c|c|}
\hline
$Z_2$ Orb. field & Scaling Dimension, $h$ & $Z_4$ parafermion field \\
\hline
\hline
$1$ & $0$ & $\Phi^0_0$ \\
\hline
$j$ & $1$ & $\Phi^0_4$ \\
\hline
$\phi^1_N$ & $3/4$ & $\Phi^0_2$ \\
\hline
$\phi^2_N$ & $3/4$ & $\Phi^0_6$ \\
\hline
$\phi_1$ & $1/12$ & $\Phi^2_2$ \\
\hline
$\phi_2$ & $1/3$ & $\Phi^2_0$ \\
\hline
$\sigma_1$ & $1/16$ & $\Phi^1_1$ \\
\hline
$\sigma_2$ & $1/16$ & $\Phi^1_{-1}$ \\
\hline
$\tau_1$ & $9/16$ & $\Phi^1_3$ \\
\hline
$\tau_2$ & $9/16$ & $\Phi^1_5$ \\
\hline
\end{tabular}
\caption{
\label{Z2fieldsN3}
Primary fields in the $Z_2$ orbifold for $N=3$, their 
scaling dimensions, and the $Z_4$ parafermion fields that they correspond to.}
\end{table}

\subsection{$O(2)$ Chern-Simons theory on a torus}

The claim of Moore and Seiberg was that $O(2)$ Chern-Simons theory at level
$2N$ corresponds to the $Z_2$ rational orbifold CFT at level $2N$. The
first step in showing this is to show that the degeneracy of this theory
on a torus is $N+7$. This is done in the following way. The classical
configuration space of pure Chern-Simons theory with gauge group $G$ consists of 
flat $G$ bundles on a torus. Flat $O(2)$ bundles can be split into two classes,
those that can be considered to be $SO(2) = U(1)$ bundles, and those that cannot.
In the first case, we simply need to take the space of states in $U(1)_{2N}$
Chern-Simons theory and keep the $Z_2$ invariant states. This leaves $N+1$
states. 

In addition to these, there are flat, twisted bundles. Flat bundles are classified
by $\text{hom }(\pi_1(M) \rightarrow G)/G$. This is the space of homomorphisms of the
fundamental group of the manifold $M$ into the gauge group $G$, modulo $G$. Let us
study the space of flat, twisted $O(2)$ bundles. We first write the gauge field 
as
\begin{align}
A_\mu = \left(\begin {array}{cc}
    a_\mu  & 0 \\
    0  & -a_\mu\\
\end{array}
\right).
\end{align}
The group is composed of $U(1)$ elements, which we write in terms of the Pauli
matrix $\sigma_3$: $e^{i \alpha \sigma_3}$.
We write the $Z_2$ element as the Pauli matrix $\sigma_1$. The $Z_2$ action
is therefore $A_\mu \rightarrow \sigma_1 A_\mu \sigma_1 = -A_\mu$. 
We can write a Lagrangian for this theory:
\begin{equation}
L = \frac{2N}{4\pi}\int_M d^2x a \partial a.
\end{equation}

We are concerned with the case where $M$ is the torus, $T^2$. $\pi_1(T^2)$
is generated by two elements, $a$ and $b$, the two non-contractible
loops of the torus. We must study the homomorphism $h : \pi_1(T^2) \rightarrow G$.  
$\pi_1(T^2)$ is an Abelian group, and since $h$ is a homomorphism, we must have:
\begin{equation}
h(\alpha a + \beta b) = \alpha \beta h(a) h(b) = \alpha \beta h(b) h(a).
\end{equation}
Suppose we are twisted in the $a$ direction only. Then, we have
\begin{equation}
h(a) = \sigma_1 e^{i\theta \sigma_3} \;\;\; h(b) = e^{i\phi \sigma_3}.
\end{equation}
Modding out by the group $O(2)$, we find that that 
$\theta \sim - \theta + 2\pi m \sim \theta + 2\alpha$, for any $\alpha$.
The first equivalence comes from modding out by the $Z_2$ element, while the 
second element comes from modding out by the $U(1)$ element. 
Similarly, $\phi \sim -\phi + 2\pi n$. $n$ and $m$ are integers. The constraint 
$h(a) h(b) = h(b) h(a)$ further implies that $\phi = 0$ or $\pi$. 
The distinct solutions to these relations are therefore that
\begin{equation}
(\theta,\phi) = (0, \pi) \mbox{ or } (0,0).
\end{equation}
A similar analysis shows that the cases in which the bundle is twisted in the 
$b$ direction only or along both $a$ and $b$ also each admit only two distinct
bundles. Therefore, there are a total of 6 distinct, twisted flat 
$O(2)$ bundles. Each corresponds to a single quantum state, for a total of $N+7$
states in the $O(2)$ Chern-Simons theory on a torus.

\subsection{$Z_2$ vortices in $O(2)$ Chern-Simons Theory}
This section is essentially an application of the analysis of $Z_2$ vortices 
in the case of $G = U(1) \times U(1) \rtimes Z_2$ to the case $G = O(2)$. 

In this case, a $Z_2$ vortex takes the gauge field to minus itself. With $n$ pairs
of $Z_2$ vortices, we again deform the manifold on which the gauge field $A_\mu$ is
defined, consider a double copy, and glue the two copies together to obtain a 
genus $g = n-1$ surface. 

The analog of eqn. (\ref{cDefn}) in this case is:
\begin{equation}
c_\mu(x,y) =  \left\{ 
  \begin{array}{ll}
    a_{\mu}(x,y) &  x \leq 0 \\
    a_{\mu}(R(x,y)) &  x \geq 0 
  \end{array} \right.
\end{equation}
and in terms of $c_\mu$ we immediately see that the action is that of a $U(1)$ Chern-Simons 
theory at level $N$:
\begin{equation}
L = \frac{2N}{4\pi}\int_{M_1} d^2x a \partial a = \frac{N}{4\pi}\int_{M} d^2x c \partial c.
\end{equation}
On a genus $g = n-1$ surface, there are $N^{n-1}$ states. But we need to project
onto the $Z_2$ invariant sector.The action of the $Z_2$ is to take $c \rightarrow -c$.
We count $(N^{n-1} +1)/2$ $Z_2$ invariant states when $N$ is odd. If $N=2$, all of the
$N^{n-1}$ states are $Z_2$ invariant.

How does this relate to the corresponding conformal field theory, which is the
$Z_2$ rational orbifold at level $2N$? Let us examine a few cases. When $N=1$, this theory is the same as
$U(1)_8$ CFT, which is abelian and which therefore should have degeneracy 1 for all $n$. 

When $N=2$, the orbifold CFT is the same as two copies of the Ising CFT. The Ising CFT has a single
non-Abelian field, $\sigma$. The space of conformal blocks corresponding to $2n$ $\sigma$ fields 
on a sphere in the Ising CFT is $2^{n-1}$, which agrees with our above analysis for $N=2$. However, a
theory with two copies of the Ising CFT would have many non-abelian fields:
\begin{align}
\sigma \otimes \mathbb{I} & \; \; \; &\sigma \otimes \psi ,
\nonumber \\
\mathbb{I} \otimes \sigma & \; \; \; &\psi \otimes \sigma ,
\nonumber \\
\sigma \otimes \sigma .
\end{align}
The space of conformal blocks corresponding to $2n$ of either $\sigma \otimes \mathbb{I}$,
$\sigma \otimes \psi$, $\mathbb{I} \otimes \sigma$, or $\psi \otimes \sigma$
will have dimension $2^{n-1}$. However, the dimension of the space of conformal blocks  corresponding
to $2n$ $\sigma \otimes \sigma$ fields will be different. Thus $Z_2$ vortices in the $O(2)$ Chern-Simons
with $N = 2$ are closely related to the fields $\sigma \otimes \mathbb{I}$,
$\sigma \otimes \psi$, $\mathbb{I} \otimes \sigma$, and $\psi \otimes \sigma$. 

When $N = 3$, the orbifold CFT is dual to the $Z_4$ parafermion CFT of Zamolodchikov and Fateev. 
We expect the $Z_2$ vortices to correspond to the $\Phi^1_1$ fields, and in fact we obtain the correct
number of states in the presence of $n$ pairs of $Z_2$ vortices, as discussed earlier.


\end{document}